%% file: main.tex
\documentclass[a4paper,fleqn]{cas-sc}

\usepackage[numbers]{natbib}
\usepackage{algorithmic}
\usepackage[linesnumbered,ruled,vlined]{algorithm2e}
\usepackage[flushleft]{threeparttable}
\usepackage{subcaption}

\def\tsc#1{\csdef{#1}{\textsc{\lowercase{#1}}\xspace}}
\tsc{WGM}
\tsc{QE}
\tsc{EP}
\tsc{PMS}
\tsc{BEC}
\tsc{DE}
\newtheorem{theorem}{Theorem}

\newdefinition{rmk}{Remark}
\newdefinition{definition}{Definition}
\newtheorem{observation}{Observation}
\newproof{pf}{Proof}
\newproof{pot}{Proof of Theorem \ref{thm2}}

\newcommand{\tc}{\boldsymbol{\sigma}}
\let\oldnl\nl
\newcommand{\nonl}{\renewcommand{\nl}{\let\nl\oldnl}}
\SetKwRepeat{Do}{do}{while}%

\begin{document}
\let\WriteBookmarks\relax
\def\floatpagepagefraction{1}
\def\textpagefraction{.001}
\shorttitle{Constrained Detecting Arrays: Mathematical Structures for Fault Identification in Combinatorial Interaction Testing}
\shortauthors{H. Jin, C. Shi and T. Tsuchiya}

\title [mode = title]{Constrained Detecting Arrays: Mathematical Structures for Fault Identification in Combinatorial Interaction Testing}



\author[1]{Hao Jin}
\cormark[1]
\ead{k-kou@ist.osaka-u.ac.jp}


\author[2]{Ce Shi}
\ead{shice060@lixin.edu.cn}

\author[1]{Tatsuhiro Tsuchiya}
\ead{t-tutiya@ist.osaka-u.ac.jp}


\address[1]{Osaka University, Yamadaoka 1-5, Suita City, Osaka, Japan}
\address[2]{Shanghai Lixin University of Accounting and Finance, 2800 Wenxiang Road, Shanghai, China}





\begin{abstract}
    \textbf{}\textbf{Context: }Detecting arrays are mathematical structures aimed at fault identification in combinatorial interaction testing. However, they cannot be directly applied to systems that have constraints among test parameters. Such constraints are prevalent in real-world systems. \\
    \textbf{Objectives: }This paper proposes Constrained Detecting Arrays~(CDAs), an extension of detecting arrays, which can be used for systems with constraints.\\
    \textbf{Methods: }The paper examines the properties and capabilities of CDAs with rigorous arguments. The paper also proposes two algorithms for constructing CDAs: One is aimed at generating minimum CDAs and the other is a heuristic algorithm aimed at fast generation of CDAs. The algorithms are evaluated through experiments using a benchmark dataset. \\
    \textbf{Results: }Experimental results show that the first algorithm can generate minimum CDAs if a sufficiently long generation time is allowed, and the second algorithm can generate minimum or near-minimum CDAs in a reasonable time.\\
    \textbf{Conclusion: }CDAs enhance detecting arrays to be applied to systems with constraints. The two proposed algorithms have different advantages with respect to the array size and generation time.
\end{abstract}




\maketitle

\input{section1.tex} 
\input{section2.tex} 
\input{section3.tex} 
\input{section4.tex} 
\input{section5.tex} 
\input{threats.tex} 

\input{section6.tex} 
\input{section7.tex} 

\bibliographystyle{model1-num-names}

\bibliography{cas-refs}





\end{document}

%% file: section1.tex
\section{Introduction}\label{sec:Intro}

\emph{Combinatorial Interaction Testing~(CIT)} is a testing approach that aims to exercise interactions among test parameters.
The basic strategy of CIT is to test all interactions among a specified number~(usually a small integer such as 2 or 3) of parameters.
Empirical results suggest that it is sufficient to only test those interactions involving a small number of parameters to reveal most of the latent faults~\cite{kuhn_introductionCT2013,kuhn_faultinteractions2004}.
Using CIT can cut off testing cost significantly when compared to exhaustive testing. 
The test suites used in CIT are usually modeled as \emph{arrays} where each row represents a test case and each column corresponds to each test parameter. 
The most typical class of arrays used for CIT is $t$-\textit{way Covering Arrays}~($t$-CAs). 
In a $t$-CA every interaction involving $t$ parameters appears in at least one test case; thus the use of a $t$-CA ensures exercising all $t$-way interactions.

There are many directions to expand the capability of CIT.
One of the directions is to add fault localization capability to test suites.
$(d,t)$-\textit{Locating Arrays}~(LAs) and $(d,t)$-\textit{Detecting Arrays}~(DAs) proposed in~\cite{colbourn_locatingarray2008} represent test suites that can not only detect but also identify faulty interactions.
The integers $d$ and $t$ are predefined parameters: $d$ represents the number of faulty interactions that can be identified and $t$ represents the number of parameters involved in the faulty interactions. 
LAs and DAs add this capability to CAs at the cost of an increased number of test cases.
 

Another direction of expanding CIT is to incorporate constraints.
Real-world systems usually have constraints on the input space.
These constraints are originated from, for example, user-defined requirements or running environment restrictions.
In order to test systems with constraints correctly, proper handling of the constraints is necessary.
For example, all test cases must satisfy the constraints.
In addition, constraints may make some interactions no longer testable.
These \emph{invalid} interactions require additional handling. 
\emph{Constrained Covering Arrays~(CCAs)} are an extension of CAs in which such constraints are incorporated into the definition.  
Many previous studies on CIT have tackled the problem of generating CCAs of small sizes~\cite{lin_TCA2015,Shiba:2004,IPOG}.

In~\cite{JIN2020110771} we proposed the concept of \emph{Constrained Locating Arrays~(CLAs)} which extends LAs by incorporating constraints. 
CLAs inherit basic properties of LAs and at the same time can be used as test suites for systems that have constraints on the input space.  
In this paper, we further develop this line of research:
We propose a new mathematical structure called \textit{Constrained Detecting Arrays}~(CDAs). 
As the name suggests, CDAs extend DAs by incorporating constraints. 
Compared to LAs, DAs allow more accurate localization of faulty interactions. 
Specifically, DAs prevent faulty interactions from being erroneously identified as non-faulty, even when there are more faulty interactions than assumed.   
CDAs extend DAs to adapt to constraints while inheriting the good property of DAs. 


The remainder of the paper is organized as follows.
In Section~\ref{sec:pre} we explain CAs, LAs, and DAs, and then describe CCAs and CLAs with examples.
In Section~\ref{sec:cda} we introduce the notion of CDAs. We also provide some basic properties of CDAs. 
In Section~\ref{sec:algo} we propose two algorithms for constructing CDAs: One leverages a solver for satisfiability problems and the other is a fast heuristic algorithm. 
The results of experiments using these algorithms are shown in Section~\ref{sec:exp}.
In Section~\ref{sec:threats} we discuss threats to validity.
In Section~\ref{sec:relatedwork} we summarize related work.
Finally, we conclude this paper in Section~\ref{sec:con}.

%% file: section2.tex
\section{Preliminaries}\label{sec:pre}

\subsection{SUT models and basic notions of CIT}

A \textit{System Under Test}~(SUT) is modeled as a 3-tuple $\mathcal{M}=\{\mathcal{F},\mathcal{S},\phi\}$.
$\mathcal{F}=\{F_1,F_2,\dots,F_k\}$ is a set of parameters in the system where $k$ is the total number of these parameters. 
$\mathcal{S}=\{S_1,S_2,\dots,S_k\}$ is a set of domains for all the parameters in $\mathcal{F}$, where $S_i$ is the domain for the parameter $F_i$.
Each domain $S_i$ consists of two or more integers ranging from 0 to $|S_i|-1$, i.e., $S_i=\{0,1,\dots,|S_i|-1\}$.
Different values of $S_i$ represent different values for the parameter~$F_i$.  
$\phi: S_1\times S_2\times\dots\times S_k\rightarrow\{true,false\}$ is a mapping representing the system constraints.
We assume that $\phi(\sigma) = true$ for some $\sigma$.

\begin{table}
    \centering
    \caption{SUT: an online shopping mobile application~\cite{HuESE2020}}
    \label{tab:example}
    \input{new_sut.tex}
\end{table}

A \emph{test case} is an element of $S_1\times S_2\times\dots\times S_k$; that is, a test case is a $k$-tuple 
$\langle \sigma_1,\dots,\sigma_i,\dots,\sigma_k\rangle$ such that $\sigma_i \in S_i$. 
A \emph{test suite} is a set of test cases. 
We view a test suite as an $N\times k$ \emph{array} where each row represents a test case and $N$ is the number of test cases. 
The \emph{size} of a test suite (i.e., array) is the number of test cases (rows) in it. 

An \emph{interaction} is a set of parameter-value pairs such that no parameters are overlapped. 
The \emph{strength} of an interaction is the number of parameter-value pairs in the interaction. 
That is, $\{(F_{i_1}, \sigma_1), \dots, (F_{i_t}, \sigma_t)\}$ is an interaction of strength~$t$  
if and only if (iff) $F_{i_j} \neq F_{i_k}$ for any $i_j, i_k \ (i_j\neq i_k)$ and $\sigma_{j} \in S_{i_j}$ for all $i_j \in \{ i_1, \dots, i_t \}$. 
We let $\curlywedge$, instead of $\emptyset$, denote the interaction of strength~0 which is an empty set. 
We say that an interaction is $t$-way iff the strength is $t$. 

An interaction $T_1$ \emph{covers} another interaction $T_2$ iff $T_2 \subseteq T_1$. 
We say that a set $\mathcal{T}$ of interactions are \emph{independent} iff $T_1 \not \subseteq T_2$ for any $T_1, T_2 \in \mathcal{T}, T_1\neq T_2$. 

We also say that a test case covers an interaction iff the value matches between the test case and the interaction on every parameter involved in the interaction.
Formally, $\tc=\langle\sigma_1,\dots,\sigma_i,\dots,\sigma_k\rangle$ covers an interaction 
$T = \{(F_{i_1}, \sigma'_{1}), \dots, (F_{i_t}, \sigma'_{t})\}$ iff  
$\sigma_{i_j} = \sigma'_{j}$ for all $j \in \{i_1, \dots, i_t\}$. 
Given an array (i.e., test suite) $A$, we express the set of rows (i.e., test cases) that cover the interaction $T$ as $\rho_A(T)$. 
Also we let $\rho_A(\mathcal{T})=\bigcup_{T\in\mathcal{T}}\rho_A(T)$. 
In words, $\rho_A(\mathcal{T})$ is the set of rows that cover at least one interaction in $\mathcal{T}$. 
Note that $\rho_A(\emptyset) = \emptyset$ and $\rho_A(\curlywedge) = A$.

A test case $\tc$ is \textit{valid} iff it satisfies the constraints $\phi$, i.e., 
$\phi(\tc) = true$; otherwise, \textit{invalid}.
The set of all valid test cases is denoted as $\mathcal{R}\ (\subseteq S_1\times S_2\times\dots\times S_k)$. 
Hence $\mathcal{R}$ is also regarded as the \emph{exhaustive test suite} 
which consists of all valid test cases. 

The valid/invalid distinction also applies to interactions: 
Interactions that no valid test cases can cover are \textit{invalid}; the other interactions, i.e., those that are covered by at least one valid test case are \textit{valid}. 
We let $\mathcal{I}_t$ and $\mathcal{VI}_t$ denote the set of all $t$-way interactions and the set of all valid $t$-way interactions, respectively. 
Similarly we let $\overline{\mathcal{I}_t}$ and $\overline{\mathcal{VI}_t}$ be the set of all interactions of strength at most $t$ and 
the set of all valid interactions of strength at most $t$. 

A valid interaction is either \emph{faulty} or \emph{non-faulty}. 
The outcome of execution of a valid test case is either \textsc{Pass} or \textsc{Fail}.
The outcome is \textsc{Fail} iff the test case covers at least one faulty interaction; the outcome is \textsc{Pass} otherwise.
The test outcome of an array is the collection of the outcomes of all rows. 

Table~\ref{tab:example} shows an SUT model which represents an online shopping mobile application.
This example, which is a modification of one in~\cite{HuESE2020}, serves as a running example throughout the paper. 
This model is formally represented as $\mathcal{M}=\{\mathcal{F},\mathcal{S},\phi\}$ where 
$\mathcal{F}=\{F_1,F_2,F_3,F_4\}$, $\mathcal{S}=\{S_1,S_2,S_3, S_4\}$, 
$S_1=S_3= \{0,1,2\}$, $S_2=\{0,1\}$, $S_4=\{0,1,2,3\}$ and 
$\phi = \phi_1 \land \phi_2 = (F_2 = 1 \Rightarrow F_3 \neq 0) \land (F_4 = 3 \Rightarrow (F_2 = 0 \land F_3 = 0))$.
An example of a valid test case is $\langle 0, 0, 0, 0\rangle$, representing 
$\langle$\$50, Domestic, Same-Day Delivery, Visa$\rangle$. 
On the other hand, $\langle 0, 1, 0, 0\rangle$ (i.e., $\langle$\$50, International, Same-Day Delivery, Visa$\rangle$)
is an invalid test case. Invalid interactions include, for example, $\{(F_2, 1), (F_3, 0)\}$, 
$\{(F_2, 1), (F_4, 3)\}$, etc. 



\subsection{Covering arrays, locating arrays, and detecting arrays}

\textit{Covering arrays}~(CAs), \textit{locating arrays}~(LAs), and \textit{detecting arrays}~(DAs) are mathematical structures that can be implemented as test suites. 
They are usually used to detect or locate faulty interactions for unconstrained SUTs~(i.e., $\phi=\textit{true}$). 

A $t$-way covering array, $t$-CA for short, is defined as follows:
\begin{equation*}
  t\textrm{-CA}\qquad\forall T\in\mathcal{I}_t:\rho_A(T)\neq\emptyset
\end{equation*}

The condition requires that all interactions $T$ in $\mathcal{I}_t$ be covered by at least one row in the array. 
In other words, when the test cases in $A$ are executed, all $t$-way interactions are examined or executed at least once. 
This condition is sufficient to reveal the existence of a $t$-way faulty interaction; 
but it is generally not possible to identify the faulty interaction from the test outcome. 
Figure~\ref{fig:2ca} shows a 2-CA for the running example. 

On the other hand, LAs and DAs can be used to not only detect the existence of faulty interactions but also locate them. 
LAs and DAs were first proposed by Colbourn and McClary in \cite{colbourn_locatingarray2008}. 
They introduced a total of six types for both LAs and DAs according to fault locating capability. 
Two types of them exist only in extreme cases. 
The rest four types, namely,  $(d,t)$-, $(\overline{d},t)$-, $(d,\overline{t})$-, $(\overline{d},\overline{t})$-LA~(and DA), are as follows ($d\geq0$, $0\leq t\leq k$). 

\begin{equation*}
  \begin{aligned}
  (d,t)\textrm{-LA}\qquad&\forall\mathcal{T}_1,\mathcal{T}_2\subseteq\mathcal{I}_t\textrm{ such that }|\mathcal{T}_1|=|\mathcal{T}_2|=d:\\
  &\rho_A(\mathcal{T}_1)=\rho_A(\mathcal{T}_2)\Leftrightarrow\mathcal{T}_1=\mathcal{T}_2\\
  (\overline{d},t)\textrm{-LA}\qquad&\forall\mathcal{T}_1,\mathcal{T}_2\subseteq\mathcal{I}_t\textrm{ such that }0\leq|\mathcal{T}_1|\leq d, 0\leq|\mathcal{T}_2|\leq d:\\
  &\rho_A(\mathcal{T}_1)=\rho_A(\mathcal{T}_2)\Leftrightarrow\mathcal{T}_1=\mathcal{T}_2
  \end{aligned}
\end{equation*}

\begin{equation*}
  \begin{aligned}
  (d,\overline{t})\textrm{-LA}\qquad&\forall\mathcal{T}_1,\mathcal{T}_2\subseteq\overline{\mathcal{I}_t}\textrm{ such that }|\mathcal{T}_1|=|\mathcal{T}_2|=d\textrm{ and }\mathcal{T}_1, \mathcal{T}_2\textrm{ are independent}:\\
  &\rho_A(\mathcal{T}_1)=\rho_A(\mathcal{T}_2)\Leftrightarrow\mathcal{T}_1=\mathcal{T}_2\\
  (\overline{d},\overline{t})\textrm{-LA}\qquad&\forall\mathcal{T}_1,\mathcal{T}_2\subseteq\overline{\mathcal{I}_t}\textrm{ such that }0\leq|\mathcal{T}_1|\leq d, 0\leq|\mathcal{T}_2|\leq d\textrm{ and }\mathcal{T}_1, \mathcal{T}_2\textrm{ are independent}:\\
  &\rho_A(\mathcal{T}_1)=\rho_A(\mathcal{T}_2)\Leftrightarrow\mathcal{T}_1=\mathcal{T}_2
  \end{aligned}
\end{equation*}

\begin{equation*}
  \begin{aligned}
  (d,t)\textrm{-DA}\qquad&\forall\mathcal{T}\subseteq\mathcal{I}_t\textrm{ such that }|\mathcal{T}|=d,\forall T\in\mathcal{I}_t:\\
  &\rho_A(T)\subseteq\rho_A(\mathcal{T})\Leftrightarrow T\in\mathcal{T}\\
  (\overline{d},t)\textrm{-DA}\qquad&\forall\mathcal{T}\subseteq\mathcal{I}_t\textrm{ such that }0\leq|\mathcal{T}|\leq d,\forall T\in\mathcal{I}_t:\\
  &\rho_A(T)\subseteq\rho_A(\mathcal{T})\Leftrightarrow T\in\mathcal{T}
  \end{aligned}
\end{equation*}

\begin{equation*}
  \begin{aligned}
  (d,\overline{t})\textrm{-DA}\qquad&\forall\mathcal{T}\subseteq\overline{\mathcal{I}_t}\textrm{ such that }|\mathcal{T}|=d,\forall T\in\overline{\mathcal{I}_t}\textrm{ and }\mathcal{T}\cup\{T\}\textrm{ is independent}:\\
  &\rho_A(T)\subseteq\rho_A(\mathcal{T})\Leftrightarrow T\in\mathcal{T}\\
  (\overline{d},\overline{t})\textrm{-DA}\qquad&\forall\mathcal{T}\subseteq\overline{\mathcal{I}_t}\textrm{ such that }0\leq|\mathcal{T}|\leq d,\forall T\in\overline{\mathcal{I}_t}\textrm{ and }\mathcal{T}\cup\{T\}\textrm{ is independent}:\\
  &\rho_A(T)\subseteq\rho_A(\mathcal{T})\Leftrightarrow T\in\mathcal{T}
  \end{aligned}
\end{equation*}

The parameter $d$ of these arrays stands for the number of faulty interactions that the array can correctly locate, while $t$ represents the strength of the target interactions. 
Writing $\overline{d}$ or $\overline{t}$ in place of $d$ or $t$ means that the array permits at most of $d$ faulty interactions or strength at most $t$. 
For example, a $(1,2)$-LA~(or DA) can locate one 2-way faulty interaction, while 
a $(\overline{2},\overline{2})$-LA~(or DA) can locate at most two faulty interactions that have strength not greater than 2.  
The reason why when dealing with $\overline{\mathcal{I}_t}$ it is required that $\mathcal{T}_1, \mathcal{T}_2$ or $\mathcal{T} \cup \{T\}$ be independent is that if there are $T_1, T_2 \in \overline{\mathcal{I}_t}$ such that $T_1 \subset T_2$, whether $T_2$ is faulty or not cannot be determined when $T_1$ is faulty.
Figure~\ref{fig:12la} and Figure~\ref{fig:12da} show a $(1,2)$-LA and a $(1,2)$-DA for the running SUT. 

In~\cite{colbourn_locatingarray2008} it is proved that 
a $(d, t)$-DA is a $(d, t)$-LA and a $(\overline{d}, t)$-LA and that a $(\overline{d}, t)$-LA is a $(d-1, t)$-DA (Lemma~7.1).
It is also proved that a $(d, t)$-DA is equivalent to a $(\overline{d}, t)$-DA and that 
a $(d, \overline{t})$-DA is equivalent to a $(\overline{d}, \overline{t})$-DA (Lemma~7.2).  
We will later provide theorems for CDAs, namely Theorems~\ref{theo:equiv3dbar} and \ref{theo:equiv3}, that are parallel to these lemmas.




\begin{figure}
  \caption{CA, LA, and DA for the running example~(constraint ignored)}
  \label{fig:arraysnoconstraints}
  \begin{minipage}[t]{0.3\textwidth}
    \subcaption{$2$-CA}
    \label{fig:2ca}
    \centering
    \input{2ca.tex}
  \end{minipage}
  \hfill
  \begin{minipage}[t]{0.3\textwidth}
    \subcaption{(1,2)-LA}
    \label{fig:12la}
    \centering
    \input{12la.tex}
  \end{minipage}
  \hfill
  \begin{minipage}[t]{0.3\textwidth}
    \subcaption{(1,2)-DA}
    \label{fig:12da}
    \centering
    \input{12da.tex}
  \end{minipage}
\end{figure}

How to identify faulty interactions using these arrays is the same as their constrained versions, namely, CLAs and CDAs. 

\subsection{Constrained versions of covering arrays and locating arrays}

CAs, LAs, and DAs do not take constraints into account. 
However real-world systems usually have complicated constraints that must be satisfied by all test cases. 

\subsubsection{Constrained covering arrays}

Constrained Covering Arrays~(CCAs) are the constrained version of CAs. 
CCAs are the most common form of test suites used in CIT: 
Most of test generation tools for CIT are in effect generators of CCAs. 
\begin{definition}[CCA]
    An array $A$ that consists of valid test cases is a $t$-CCA iff the following condition holds. 
    \begin{equation*}
        t\textrm{-CCA}\qquad\forall T\in\mathcal{VI}_t:\rho_A(T)\neq\emptyset
    \end{equation*}
\end{definition}
The definition of CCAs requires that all valid $t$-way interactions be covered by at least one test case in the test suite. 
This condition implies that  every valid interaction of strength $< t$ is covered by at least one test case. 
That is, a $t$-CCA is also a $(t-1)$-CCA when $t>0$. 
Thus, a $t$-CCA can also be defined as follows. 
\begin{equation*}
    t\textrm{-CCA}\qquad\forall T\in\overline{\mathcal{VI}_t}:\rho_A(T)\neq\emptyset
\end{equation*}

Figure~\ref{fig:2cca} shows a 2-CCA for the running example. 
All four invalid 2-way interactions which violate the constraints are listed as follows. 
 \begin{equation*}
     \begin{aligned}
        \{(F_2,1),(F_3,0)\}\qquad\{(F_2,1),(F_4,3)\}\\
        \{(F_3,2),(F_4,3)\}\qquad\{(F_3,1),(F_4,3)\}
     \end{aligned}
 \end{equation*}
 It is easy to observe that none of the invalid interactions appears in any rows in Figure~\ref{fig:2cca}. 


\begin{figure}
  \caption{CCA and CLA for the running example}
  \label{fig:arrayswithconstraints}
  \begin{minipage}[t]{0.4\textwidth}
    \subcaption{$2$-CCA}
    \label{fig:2cca}
    \centering
    \input{2cca.tex}
  \end{minipage}
  \hfill
  \begin{minipage}[t]{0.4\textwidth}
    \subcaption{(1,2)-CLA}
    \label{fig:12cla}
    \centering
    \input{12cla.tex}
  \end{minipage}
\end{figure}

\subsubsection{Constrained locating arrays}

LAs allow us to identify the set of faulty interactions using the test outcome.
This becomes possible because for any LA $A$, $\rho_A(\cdot)$ injectively maps an interaction set to a test outcome 
(i.e., $\rho_A(\mathcal{T}_1) = \rho_A(\mathcal{T}_2) \Rightarrow \mathcal{T}_1 = \mathcal{T}_2$). 
Since a test outcome corresponds to at most one interaction set, the set of faulty interactions can be uniquely inferred from the test outcome. 

When incorporating constraints into LAs, it is necessary to handle the situation where constraints may prevent some sets of faulty interactions from being identified. 
For the running example, when 
either one of the two interaction sets shown below is faulty, 
it is impossible to determine which is indeed faulty.
\begin{equation*}
    \mathcal{T}_1=\{\{(F_2,0),(F_4,3)\}\} \quad \mathcal{T}_2=\{\{(F_3,0),(F_4,3)\}\}
\end{equation*}
This is because the two interactions always appear simultaneously in any valid test case and thus no valid test case exists that yields different outcomes for the two faulty interaction sets. 
We say that  two sets of valid interactions, $\mathcal{T}$ and $\mathcal{T}'$, are not \emph{distinguishable} or \emph{indistinguishable} iff $\rho_A(\mathcal{T}) = \rho_A(\mathcal{T}')$ for any array $A$ that consists of valid tests. 

The definition of CLAs adapts LAs to the presence of indistinguishable interaction sets by exempting them from fault identification. 
\begin{definition}[CLA]
    Let $d\geq0$ and $0\leq t\leq k$. 
An array $A$ that consists of valid tests is a $(d,t)$-, $(\overline{d},t)$-, $(d,\overline{t})$- or $(\overline{d},\overline{t})$-CLA iff the corresponding condition shown below holds. 
    \begin{equation*}
        \begin{aligned}
            (d,t)\textrm{-CLA}\qquad&\forall\mathcal{T}_1,\mathcal{T}_2\subseteq\mathcal{VI}_t\textrm{ such that }|\mathcal{T}_1|=|\mathcal{T}_2|=d\textrm{ and }\mathcal{T}_1,\mathcal{T}_2\textrm{ are distinguishable}:\\
            &\rho_A(\mathcal{T}_1)\neq\rho_A(\mathcal{T}_2)\\
            (\overline{d},t)\textrm{-CLA}\qquad&\forall\mathcal{T}_1,\mathcal{T}_2\subseteq\mathcal{VI}_t\textrm{ such that }0\leq|\mathcal{T}_1|\leq d,0\leq|\mathcal{T}_2|\leq d\textrm{ and }\mathcal{T}_1,\mathcal{T}_2\textrm{ are distinguishable}:\\
            &\rho_A(\mathcal{T}_1)\neq\rho_A(\mathcal{T}_2)\\
            (d,\overline{t})\textrm{-CLA}\qquad&\forall\mathcal{T}_1,\mathcal{T}_2\subseteq\overline{\mathcal{VI}_t}\textrm{ such that }|\mathcal{T}_1|=|\mathcal{T}_2|=d\textrm{ and }\mathcal{T}_1,\mathcal{T}_2\textrm{ are independent and distinguishable}:\\
            &\rho_A(\mathcal{T}_1)\neq\rho_A(\mathcal{T}_2)\\
            (\overline{d},\overline{t})\textrm{-CLA}\qquad&\forall\mathcal{T}_1,\mathcal{T}_2\subseteq\overline{\mathcal{VI}_t}\textrm{ such that }0\leq|\mathcal{T}_1|\leq d,0\leq|\mathcal{T}_2|\leq d\textrm{ and }\mathcal{T}_1,\mathcal{T}_2\textrm{ are independent and distinguishable}:\\
            &\rho_A(\mathcal{T}_1)\neq\rho_A(\mathcal{T}_2)
        \end{aligned}
    \end{equation*}
\end{definition}

A (1,2)-CLA for the running example is shown in Figure~\ref{fig:12cla}. 
From the array, one can see that none of the invalid interactions shown above appears in any rows of the CLA. 
In addition, for any pair of 2-way interactions except the above indistinguishable pair, the rows that cover one of them are different from those that cover the other.
The exception is the rows where the indistinguishable pair of interaction sets appear, namely $\boldsymbol{\sigma}_2$, $\boldsymbol{\sigma}_7$ and $\boldsymbol{\sigma}_{13}$. 
This occurs because of the second constraint $\phi_2$ in the SUT. 
$\phi_2$ enforces all test cases that have $(F_4,3)$ to contain $(F_2,0)$ and $(F_3,0)$ at the same time. 
Thus, the interaction sets $\mathcal{T}_1=\{\{(F_2,0),(F_4,3)\}\}$ and $\mathcal{T}_2=\{\{(F_3,0),(F_4,3)\}\}$ appear in the same test cases as long as the test cases are valid. 



%% file: new_sut.tex
\begin{tabular}{c|llll}
\hline 
$\mathcal{F}$ & $F_1$ (Total Price) & $F_2$ (Shipping Address) & $F_3$ (Shipping Method) & $F_4$ (Payment Method)\tabularnewline
\hline 
\multirow{4}{*}{$\mathcal{S}$} & 0: \$50 & 0: Domestic & 0: Same-Day Delivery & 0: Visa\tabularnewline
 & 1: \$500 & 1: International & 1: 2-Day Delivery & 1: Mastercard\tabularnewline
 & 2: \$1000 & -- -- & 2: 7-Day Delivery & 2: Paypal\tabularnewline
 & -- -- & -- -- & -- -- & 3: Gift Card\tabularnewline
\hline 
\multirow{2}{*}{$\phi$} & \multicolumn{4}{l}{$\phi_1:\textrm{Shipping Address}=\textit{International}\Rightarrow\textrm{Shipping Method}\neq\textit{Same-Day Delivery}$}\tabularnewline
 & \multicolumn{4}{l}{$\phi_2:\textrm{Payment}=\textit{Gift Card}\Rightarrow\textrm{Shipping Address}=\textit{Domestic}\land\textrm{Shipping Method}=\textit{Same-Day Delivery}$}\tabularnewline
\hline 
\end{tabular}

%% file: 2ca.tex
\begin{tabular}{c|cccc}
\hline 
 & $F_1$ & $F_2$ & $F_3$ & $F_4$\tabularnewline
\hline 
$\boldsymbol\sigma_1$ & 0 & 0 & 0 & 0\tabularnewline
$\boldsymbol\sigma_2$ & 0 & 0 & 2 & 2\tabularnewline
$\boldsymbol\sigma_3$ & 0 & 1 & 0 & 3\tabularnewline
$\boldsymbol\sigma_4$ & 0 & 1 & 1 & 1\tabularnewline
$\boldsymbol\sigma_5$ & 1 & 0 & 0 & 1\tabularnewline
$\boldsymbol\sigma_6$ & 1 & 0 & 2 & 3\tabularnewline
$\boldsymbol\sigma_7$ & 1 & 1 & 0 & 2\tabularnewline
$\boldsymbol\sigma_8$ & 1 & 1 & 1 & 0\tabularnewline
$\boldsymbol\sigma_9$ & 2 & 0 & 0 & 3\tabularnewline
$\boldsymbol\sigma_{10}$ & 2 & 0 & 1 & 2\tabularnewline
$\boldsymbol\sigma_{11}$ & 2 & 0 & 2 & 0\tabularnewline
$\boldsymbol\sigma_{12}$ & 2 & 1 & 1 & 3\tabularnewline
$\boldsymbol\sigma_{13}$ & 2 & 1 & 2 & 1\tabularnewline
\hline 
\end{tabular}

%% file: 12la.tex
\begin{tabular}{c|cccc}
\hline 
 & $F_1$ & $F_2$ & $F_3$ & $F_4$\tabularnewline
\hline 
$\boldsymbol\sigma_1$ & 0 & 0 & 0 & 0\tabularnewline
$\boldsymbol\sigma_2$ & 0 & 0 & 1 & 1\tabularnewline
$\boldsymbol\sigma_3$ & 0 & 0 & 1 & 2\tabularnewline
$\boldsymbol\sigma_4$ & 0 & 1 & 0 & 3\tabularnewline
$\boldsymbol\sigma_5$ & 0 & 1 & 2 & 0\tabularnewline
$\boldsymbol\sigma_6$ & 0 & 1 & 2 & 2\tabularnewline
$\boldsymbol\sigma_7$ & 1 & 0 & 0 & 1\tabularnewline
$\boldsymbol\sigma_8$ & 1 & 0 & 1 & 0\tabularnewline
$\boldsymbol\sigma_9$ & 1 & 0 & 2 & 1\tabularnewline
$\boldsymbol\sigma_{10}$ & 1 & 1 & 0 & 0\tabularnewline
$\boldsymbol\sigma_{11}$ & 1 & 1 & 0 & 2\tabularnewline
$\boldsymbol\sigma_{12}$ & 1 & 1 & 1 & 1\tabularnewline
$\boldsymbol\sigma_{13}$ & 1 & 1 & 2 & 3\tabularnewline
$\boldsymbol\sigma_{14}$ & 2 & 0 & 0 & 2\tabularnewline
$\boldsymbol\sigma_{15}$ & 2 & 0 & 0 & 3\tabularnewline
$\boldsymbol\sigma_{16}$ & 2 & 0 & 1 & 1\tabularnewline
$\boldsymbol\sigma_{17}$ & 2 & 0 & 2 & 3\tabularnewline
$\boldsymbol\sigma_{18}$ & 2 & 1 & 0 & 0\tabularnewline
$\boldsymbol\sigma_{19}$ & 2 & 1 & 1 & 3\tabularnewline
\hline 
\end{tabular}

%% file: 12da.tex
\begin{tabular}{c|cccc}
\hline 
 & $F_1$ & $F_2$ & $F_3$ & $F_4$\tabularnewline
\hline 
$\boldsymbol\sigma_1$ & 0 & 0 & 0 & 0\tabularnewline
$\boldsymbol\sigma_2$ & 0 & 0 & 0 & 2\tabularnewline
$\boldsymbol\sigma_3$ & 0 & 0 & 1 & 1\tabularnewline
$\boldsymbol\sigma_4$ & 0 & 0 & 2 & 3\tabularnewline
$\boldsymbol\sigma_5$ & 0 & 1 & 0 & 1\tabularnewline
$\boldsymbol\sigma_6$ & 0 & 1 & 1 & 3\tabularnewline
$\boldsymbol\sigma_7$ & 0 & 1 & 2 & 0\tabularnewline
$\boldsymbol\sigma_8$ & 0 & 1 & 2 & 2\tabularnewline
$\boldsymbol\sigma_9$ & 1 & 0 & 0 & 1\tabularnewline
$\boldsymbol\sigma_{10}$ & 1 & 0 & 1 & 3\tabularnewline
$\boldsymbol\sigma_{11}$ & 1 & 0 & 2 & 0\tabularnewline
$\boldsymbol\sigma_{12}$ & 1 & 0 & 2 & 2\tabularnewline
$\boldsymbol\sigma_{13}$ & 1 & 1 & 0 & 3\tabularnewline
$\boldsymbol\sigma_{14}$ & 1 & 1 & 1 & 0\tabularnewline
$\boldsymbol\sigma_{15}$ & 1 & 1 & 1 & 2\tabularnewline
$\boldsymbol\sigma_{16}$ & 1 & 1 & 2 & 1\tabularnewline
$\boldsymbol\sigma_{17}$ & 2 & 0 & 0 & 3\tabularnewline
$\boldsymbol\sigma_{18}$ & 2 & 0 & 1 & 0\tabularnewline
$\boldsymbol\sigma_{19}$ & 2 & 0 & 1 & 2\tabularnewline
$\boldsymbol\sigma_{20}$ & 2 & 0 & 2 & 1\tabularnewline
$\boldsymbol\sigma_{21}$ & 2 & 1 & 0 & 0\tabularnewline
$\boldsymbol\sigma_{22}$ & 2 & 1 & 0 & 2\tabularnewline
$\boldsymbol\sigma_{23}$ & 2 & 1 & 1 & 1\tabularnewline
$\boldsymbol\sigma_{24}$ & 2 & 1 & 2 & 3\tabularnewline
\hline 
\end{tabular}

%% file: 2cca.tex
\begin{tabular}{c|cccc}
\hline 
 & $F_1$ & $F_2$ & $F_3$ & $F_4$\tabularnewline
\hline 
$\boldsymbol\sigma_1$ & 0 & 0 & 0 & 0\tabularnewline
$\boldsymbol\sigma_2$ & 0 & 0 & 0 & 3\tabularnewline
$\boldsymbol\sigma_3$ & 0 & 1 & 1 & 1\tabularnewline
$\boldsymbol\sigma_4$ & 0 & 1 & 2 & 2\tabularnewline
$\boldsymbol\sigma_5$ & 1 & 0 & 0 & 2\tabularnewline
$\boldsymbol\sigma_6$ & 1 & 0 & 0 & 3\tabularnewline
$\boldsymbol\sigma_7$ & 1 & 0 & 2 & 1\tabularnewline
$\boldsymbol\sigma_8$ & 1 & 1 & 1 & 0\tabularnewline
$\boldsymbol\sigma_9$ & 2 & 0 & 0 & 1\tabularnewline
$\boldsymbol\sigma_{10}$ & 2 & 0 & 0 & 3\tabularnewline
$\boldsymbol\sigma_{11}$ & 2 & 0 & 1 & 2\tabularnewline
$\boldsymbol\sigma_{12}$ & 2 & 1 & 2 & 0\tabularnewline
\hline 
\end{tabular}

%% file: 12cla.tex
\begin{tabular}{c|cccc}
\hline 
 & $F_1$ & $F_2$ & $F_3$ & $F_4$\tabularnewline
\hline 
$\boldsymbol\sigma_1$ & 0 & 0 & 0 & 0\tabularnewline
$\boldsymbol\sigma_2$ & 0 & 0 & 0 & 3\tabularnewline
$\boldsymbol\sigma_3$ & 0 & 0 & 1 & 1\tabularnewline
$\boldsymbol\sigma_4$ & 0 & 1 & 1 & 2\tabularnewline
$\boldsymbol\sigma_5$ & 0 & 1 & 2 & 0\tabularnewline
$\boldsymbol\sigma_6$ & 1 & 0 & 0 & 2\tabularnewline
$\boldsymbol\sigma_7$ & 1 & 0 & 0 & 3\tabularnewline
$\boldsymbol\sigma_8$ & 1 & 0 & 1 & 2\tabularnewline
$\boldsymbol\sigma_9$ & 1 & 1 & 1 & 1\tabularnewline
$\boldsymbol\sigma_{10}$ & 1 & 1 & 2 & 0\tabularnewline
$\boldsymbol\sigma_{11}$ & 1 & 1 & 2 & 2\tabularnewline
$\boldsymbol\sigma_{12}$ & 2 & 0 & 0 & 1\tabularnewline
$\boldsymbol\sigma_{13}$ & 2 & 0 & 0 & 3\tabularnewline
$\boldsymbol\sigma_{14}$ & 2 & 0 & 2 & 0\tabularnewline
$\boldsymbol\sigma_{15}$ & 2 & 1 & 1 & 0\tabularnewline
$\boldsymbol\sigma_{16}$ & 2 & 1 & 1 & 2\tabularnewline
$\boldsymbol\sigma_{17}$ & 2 & 1 & 2 & 1\tabularnewline
\hline 
\end{tabular}

%% file: section3.tex
\section{Constrained Detecting Arrays}\label{sec:cda}

In this section, we propose Constrained Detecting Arrays (CDAs). 
First we present the definition of CDAs; then we show how one can identify faulty interactions 
with CDAs. 
In addition, we provide some theorems that relate CDAs to CCAs and CLAs.

\subsection{Definition}

For an array $A$ to be a DA, $A$ must satisfy $\rho_A(T)\subseteq\rho_A(\mathcal{T})\Leftrightarrow T\in\mathcal{T}$ for any pair of an interaction $T$ and an interaction set $\mathcal{T}$. 
However, this may be impossible if $A$ consists only of valid test cases. 
Here we introduce the concept of \emph{masking} to capture such a situation. 

\begin{definition}[Masking]
    A set $\mathcal{T}$ of valid interactions \emph{masks} a valid interaction $T$ iff $T\not\in\mathcal{T}$ and
    \begin{equation*}
        \forall\boldsymbol{\sigma}\in\mathcal{R}:T\subseteq\boldsymbol{\sigma}\Rightarrow(\exists T'\in\mathcal{T}:T'\subseteq\boldsymbol{\sigma}). 
    \end{equation*}
    If $\mathcal{T}$ masks $T$, we write $\mathcal{T}\succ T$; 
	otherwise we write $\mathcal{T}\not\succ T$. 
By definition,  $\mathcal{T}\not\succ T$ iff  
$T \in \mathcal{T}$ or 
	\[
	\exists \boldsymbol{\sigma}\in \mathcal{R}: T\subseteq \sigma \land 
(\forall T'\in \mathcal{T}:T'\not\subseteq \sigma).
\]
\end{definition}

In words, when $\mathcal{T}$ masks $T$, $T$ always appears together with some interaction $T'$ in $\mathcal{T}$ in any valid test case $\sigma$. 
In this case, $T \not \in \mathcal{T}$ but 
$\rho_A(T) \subseteq \rho_A(\mathcal{T})$ always holds for any $A$ that meets the constraints. 
In the running example, such $T$-$\mathcal{T}$ pairs include: 
\begin{equation*}
    \begin{aligned}
        \mathcal{T}_1=\{\{(F_1,0),(F_2,0)\}\}\succ T_a=\{(F_1,0),(F_3,0)\}\qquad\mathcal{T}_1=\{\{(F_1,0),(F_2,0)\}\}\succ T_b=\{(F_1,0),(F_4,3)\}\\
        \mathcal{T}_2=\{\{(F_1,1),(F_2,0)\}\}\succ T_c=\{(F_1,1),(F_3,0)\}\qquad\mathcal{T}_3=\{\{(F_2,0),(F_3,0)\}\}\succ T_b=\{(F_1,0),(F_4,3)\}\\
        \mathcal{T}_4=\{\{(F_3,0),(F_4,3)\}\}\succ T_d=\{(F_1,2),(F_4,3)\}\qquad\mathcal{T}_4=\{\{(F_3,0),(F_4,3)\}\}\succ T_e=\{(F_2,0),(F_4,3)\}\\
        \dots\dots(\textrm{31 pairs in total})
    \end{aligned}
\end{equation*}

When $\mathcal{T}$ masks $T$, the failure caused by $T$ cannot be inherently distinguished from that caused by $\mathcal{T}$. 
The idea at the core of CDAs is to relax the condition of DAs by exempting $T$-$\mathcal{T}$ pairs such that $\mathcal{T}\succ T$ from fault identification.

\begin{definition}[CDA]
    Let $d\geq0$ and $0\leq t\leq k$. An array $A$ that consists of valid test cases or no rows is a $(d,t)$-, $(\overline{d},t)$-, $(d,\overline{t})$- or $(\overline{d},\overline{t})$-CDA iff the corresponding condition shown below holds. 
    \begin{equation*}
        \begin{aligned}
            (d,t)\textrm{-CDA}\qquad&\forall\mathcal{T}\subseteq\mathcal{VI}_t\textrm{ such that }|\mathcal{T}|=d,\forall T\in\mathcal{VI}_t:\\
            &\mathcal{T}\not\succ T\Rightarrow(T\in\mathcal{T}\Leftrightarrow\rho_A(T)\subseteq\rho_A(\mathcal{T}))\\
            (\overline{d},t)\textrm{-CDA}\qquad&\forall\mathcal{T}\subseteq\mathcal{VI}_t\textrm{ such that }0\leq|\mathcal{T}|\leq d,\forall T\in\mathcal{VI}_t:\\
            &\mathcal{T}\not\succ T\Rightarrow(T\in\mathcal{T}\Leftrightarrow\rho_A(T)\subseteq\rho_A(\mathcal{T}))\\
            (d,\overline{t})\textrm{-CDA}\qquad&\forall\mathcal{T}\subseteq\overline{\mathcal{VI}_t}\textrm{ such that }|\mathcal{T}|=d,\forall T\in\overline{\mathcal{VI}_t}\textrm{ and }\mathcal{T}\cup\{T\}\textrm{ is independent}:\\
            &\mathcal{T}\not\succ T\Rightarrow(T\in\mathcal{T}\Leftrightarrow\rho_A(T)\subseteq\rho_A(\mathcal{T}))\\
            (\overline{d},\overline{t})\textrm{-CDA}\qquad&\forall\mathcal{T}\subseteq\overline{\mathcal{VI}_t}\textrm{ such that }0\leq|\mathcal{T}|\leq d,\forall T\in\overline{\mathcal{VI}_t}\textrm{ and }\mathcal{T}\cup\{T\}\textrm{ is independent}:\\
            &\mathcal{T}\not\succ T\Rightarrow(T\in\mathcal{T}\Leftrightarrow\rho_A(T)\subseteq\rho_A(\mathcal{T}))\\
        \end{aligned}
    \end{equation*}
\end{definition}

Figure~\ref{fig:11cda}, Figure~\ref{fig:21cda}, and Figure~\ref{fig:12cda} respectively show a (1,1)-CDA, a (2,1)-CDA, and a (1,2)-CDA for the running example.
Now let us take the (1,2)-CDA in Figure~\ref{fig:12cda} and the pair of $\mathcal{T}_3=\{\{(F_2,0),(F_3,0)\}\}$ and $T_b=\{(F_1,0),(F_4,3)\}$ as examples. 
Let $A$ denote the (1,2)-CDA for now; then
$\rho_A(\mathcal{T}_3) = \{\boldsymbol{\sigma}_1$, 
$\boldsymbol{\sigma}_2$, $\boldsymbol{\sigma}_3$, $\boldsymbol{\sigma}_9$, $\boldsymbol{\sigma}_{10}$, $\boldsymbol{\sigma}_{16}$, $\boldsymbol{\sigma}_{17}$, $\boldsymbol{\sigma}_{18}$, $\boldsymbol{\sigma}_{19}\}$, 
whereas $\rho_A(T_b) = \{\boldsymbol{\sigma}_3\}$. 
Hence, $T_b \in \mathcal{T}_3 \Leftrightarrow \rho_A(T_b) \subseteq \rho_A(\mathcal{T}_3)$ does not hold. 
This is prohibited by the definition of DAs but is allowed by CDAs, because $\mathcal{T}_3 \succ T_b$, which means 
that no array can satisfy it unless the constraints are violated. 




\begin{figure}
    \caption{CDAs for the running example}
    \label{fig:cdaarrays}
    \begin{minipage}[t]{0.3\textwidth}
      \subcaption{(1,1)-CDA}
      \label{fig:11cda}
      \centering
      \input{11cda.tex}
    \end{minipage}
    \hfill
    \begin{minipage}[t]{0.3\textwidth}
      \subcaption{(2,1)-CDA}
      \label{fig:21cda}
      \centering
      \input{21cda.tex}
    \end{minipage}
    \hfill
    \begin{minipage}[t]{0.3\textwidth}
      \subcaption{(1,2)-CDA}
      \label{fig:12cda}
      \centering
      \input{12cda.tex}
    \end{minipage}
  \end{figure}

By definition, it is straightforward to see that the following observations hold. 

\begin{observation}
    A $(\overline{d},\overline{t})$-CDA is a $(\overline{d},t)$-CDA and a $(d,\overline{t})$-CDA. A $(\overline{d},t)$-CDA and a $(d,\overline{t})$-CDA are both a $(d,t)$-CDA. 
  When $d > 0$, a $(\overline{d},\overline{t})$-CDA and a $(\overline{d},t)$-CDA are a $(\overline{d-1},\overline{t})$-CDA and a $(\overline{d-1},t)$-CDA, respectively. 
  When $t > 0$, a $(\overline{d},\overline{t})$-CDA and a $(d,\overline{t})$-CDA are a $(\overline{d},\overline{t-1})$-CDA and a $(d,\overline{t-1})$-CDA, respectively.
\end{observation}

\begin{observation}
    Suppose that the SUT has no constraints, i.e., $\phi(\boldsymbol{\sigma})=true$ for all $\boldsymbol{\sigma}\in\mathcal{R}=V_1\times V_2\times\dots\times V_k$ 
 and that a $(d, t)$-DA $A$ exists. Then $A$ is a $(d, t)$-CDA. 
This also applies when $d$ or $t$ is replaced with $\overline{d}$ or $\overline{t}$, respectively. 
\end{observation}

According to the above definition, if $|\mathcal{T}|$ is very large, then $\rho_A(\mathcal{T}) = A$ for any array $A$, in which case all interactions are masked by $\mathcal{T}$.
In order to avoid such cases of no practical interest, here we introduce an upper bound, 
denoted $\tau_t$, on $d$. 
We let $\tau_t \geq 0$ be the greatest integer that satisfies the condition as follows: 
\[
\forall \mathcal{T} \subseteq \mathcal{VI}_t \mathrm{~such~that~} |\mathcal{T}| \leq \tau_t:  
\mathcal{R} - \rho_{\mathcal{R}}(\mathcal{T}) \neq \emptyset 
\]
In words, given $\tau_t$ interactions of strength $t$, there is always a test case in $\mathcal{R}$ 
in which none of the given interactions appears.
Note that $\tau_0 = 0$.


\begin{theorem}
	\label{theo:equiv2}
	For $d \leq \tau_t$,  
	a $(d,t)$-CDA is equivalent to a $(\overline{d},t)$-CDA.
\end{theorem}

\begin{pf}
Trivially a $(0,t)$-CDA is a $(\overline{0},t)$-CDA. 
	Let $A$ be a $(d,t)$-CDA such that $1\leq d \leq \tau_t$ and $t>0$. We will show that $A$ is a $(d-1,t)$-CDA.
	Let $\mathcal{T}$ and $T$ be a set of $d-1$ valid interactions of strength $t$ 
	and a $t$-way valid interaction, respectively.
	If $\mathcal{T} \succ T$ or $T \in \mathcal{T}$, then $\mathcal{T} \not \succ T \Rightarrow (T \in \mathcal{T} \Leftrightarrow \rho_A(T) \subseteq \rho_A(\mathcal{T}))$ trivially holds.
	The rest of the proof considers the case where $T \not\in \mathcal{T}$ and $\mathcal{T} \not \succ T$.
	In this case, there is some $\sigma \in \mathcal{R}$ such that $T \subseteq \sigma$ and 
	$\sigma \not\in \rho_{\mathcal{R}}(\mathcal{T})$.
	Since $|\mathcal{T}\cup \{T\}| \leq \tau_t$, 
	$\mathcal{R} - \rho_{\mathcal{R}}(\mathcal{T}\cup \{T\})$ is not empty. 
	Let $T'$ be any $t$-way interaction that appears in a test case in
	$\mathcal{R} - \rho_{\mathcal{R}}(\mathcal{T}\cup \{T\})$
	and has exactly the same $t$ parameters as $T$. 
	Note that $T$ and $T'$ cannot appear in any test case simultaneously. 
	Let $\mathcal{T}' = \mathcal{T} \cup \{T'\}$. 
	$\mathcal{T}' \not\succ T$
	since $T \subseteq \sigma$,  $\sigma \not\in \rho_{\mathcal{R}}(\mathcal{T})$, 
	and $\sigma \not\in \rho_{\mathcal{R}}(T')$.
	Because $A$ is a $(d, t)$-CDA and $\mathcal{T}' \not \succ T$, 
	$\rho_A(T) \not \subseteq \rho_A(\mathcal{T}')$.
	Hence $\rho_A(T) \not \subseteq \rho_A(\mathcal{T})$, which means 
	   $\mathcal{T} \not \succ T \Rightarrow (T \in \mathcal{T} \Leftrightarrow \rho_A(T) \subseteq \rho_A(\mathcal{T}))$.
	By induction, $A$ is a $(d',t)$-CDA for any $0 \leq d' \leq d$ and thus is a $(\overline{d}, t)$-CDA. \qed
\end{pf}

A similar argument applies to $(d, \overline{t})$-CDAs and $(\overline{d}, \overline{t})$-CDAs.
\begin{theorem}
	\label{theo:equiv3dbar}
	For $d = t = 0$ or $d \leq \tau_1$ and $t > 0$,  
	a $(d,\overline{t})$-CDA is equivalent to a $(\overline{d},\overline{t})$-CDA. 
\end{theorem}

\begin{pf}
Trivially $(0, \overline{t})$-CDA is a $(\overline{0}, \overline{t})$-CDA.
	Let $A$ be a $(d,\overline{t})$-CDA such that $1\leq d \leq \tau_t$ and $t>0$. 
Below we will show that $A$ is a $(d-1,\overline{t})$-CDA.
	Let $\mathcal{T} \subseteq \overline{\mathcal{VI}_t}$ such that $\mathcal{T}$ is independent and $|\mathcal{T}|=d-1$. 
Let $T$ be a valid interaction of strength at most $t$.  
	If $\mathcal{T} \succ T$ or $T \in \mathcal{T}$, then $\mathcal{T} \not \succ T \Rightarrow (T \in \mathcal{T} \Leftrightarrow \rho_A(T) \subseteq \rho_A(\mathcal{T}))$ trivially holds.

	Consider the remaining case where $\mathcal{T} \not \succ T$ and $T \not\in \mathcal{T}$.
	In this case, there is some $\sigma \in \mathcal{R}$ such that $T \subseteq \sigma$ and 
	$\sigma \not\in \rho_{\mathcal{R}}(\mathcal{T})$.
%

Case $|T| > 0$.
	Since $|\mathcal{T}\cup \{ T\}| = d \leq \tau_1$
and every interaction in $\mathcal{T}\cup \{ T\}$ has strength at least one,  
	$\mathcal{R} - \rho_{\mathcal{R}}(\mathcal{T}\cup \{ T\})$ is not empty. 
	Let $T'$ be an interaction of strength $t$ that appears in a test case in
	$\mathcal{R} - \rho_{\mathcal{R}}(\mathcal{T}\cup \{T\})$ 
	and has a different value on at least one parameter from $T$. 
	Let $\mathcal{T}' = \mathcal{T} \cup \{T'\}$. 
	$\mathcal{T}'$ is independent because $\mathcal{T}$ is independent 
	and $\hat T \not\subset T'$ for any $\hat{T} \in \mathcal{T}$.
    (Note that if $\hat T \subset T'$, $\hat T$ would occur in the test case with $T'$.) 
Also
	$\mathcal{T}' \not\succ T$
	since $T \subseteq \sigma$,  $\sigma \not\in \rho_{\mathcal{R}}(\mathcal{T})$, 
	and $\sigma \not\in \rho_{\mathcal{R}}(T')$.
	Because $A$ is a $(d, \overline{t})$-CDA, $\mathcal{T}'$ is independent, and $\mathcal{T}' \not \succ T$, 
	we have $\rho_A(T) \not \subseteq \rho_A(\mathcal{T}')$.
	Hence $\rho_A(T) \not \subseteq \rho_A(\mathcal{T})$.

Case $T = \curlywedge$ (i.e., $|T| = 0$).
As $\curlywedge \not\in\mathcal{T}$, every interaction in $\mathcal{T}$ has strength at least one. 
Since $|\mathcal{T}| = d - 1 < \tau_1$,    
	$\mathcal{R} - \rho_{\mathcal{R}}(\mathcal{T})$ is not empty. 
	Let $T'$ be any $t$-way interaction that appears in a test case in
	$\mathcal{R} - \rho_{\mathcal{R}}(\mathcal{T})$. 
Let $\mathcal{T}' = \mathcal{T} \cup \{T'\}$.
Because of the same argument as in the case $|T|>0$, 
$\mathcal{T}'$ is independent and $\mathcal{T}' \not \succ T$ and thus  
	$\rho_A(T) \not \subseteq \rho_A(\mathcal{T}')$.
	Hence $\rho_A(T) \not \subseteq \rho_A(\mathcal{T})$. 

As a result,  
	   $\mathcal{T} \not \succ T \Rightarrow (T \in \mathcal{T} \Leftrightarrow \rho_A(T) \subseteq \rho_A(\mathcal{T}))$.
By induction, $A$ is a $(d',\overline{t})$-CDA for any $0 \leq d' \leq d$ and thus is a $(\overline{d}, \overline{t})$-CDA. \qed
\end{pf}

\begin{figure}[t]
    \caption{The relationships among CDA variants~(when marked with $+$, one requires that $d \leq \tau_t $; when marked with $\dag$, $d \leq \tau_1$)}\label{fig:cdarelation}
    \centering
    \begin{tabular}{ccc}
        $(d,\overline{t})\textrm{-CDA}$ & $\Rightarrow$ & $(d,t)\textrm{-CDA}$ \\
        $\Uparrow$$\Downarrow^+$&  &$\Uparrow$$\Downarrow^\dag$ \\
        $(\overline{d},\overline{t})\textrm{-CDA}$ & $\Rightarrow$ & $(\overline{d},t)\textrm{-CDA}$ 
    \end{tabular}
\end{figure}
These theorems lead to the relationships illustrated in Figure~\ref{fig:cdarelation}.
Because of these results, we henceforth focus on  
$(\overline{d}, t)$-CDAs and $(\overline{d}, \overline{t})$-CDAs.

\begin{theorem}\label{thr:cda2cca}
	A $(\overline{d}, t)$-CDA is a $t$-CCA.  
	A $(\overline{d}, \overline{t})$-CDA is a $t$-CCA.  
\end{theorem}

\begin{pf}
	Let $T \in \mathcal{VI}_t$.
	Let $A$ be a $(\overline{d},t)$-CDA or a $(\overline{d}, \overline{t})$-CDA. 
	Then $T \not\succ T \Rightarrow (T \in \mathcal{T} \Leftrightarrow \rho_A(T) \subseteq \rho_A(\mathcal{T}))$ for any $\mathcal{T} \subseteq \mathcal{VI}_t$ such that $|\mathcal{T}| \leq d$. 
	If $|\mathcal{T}| = 0$, then $\mathcal{T} = \emptyset$, in which case $\mathcal{T}\not\succ T$, $T \not \in \mathcal{T}$, and $\rho_A(\mathcal{T}) = \emptyset$.
	Hence $\rho_A(T) \not= \emptyset$. \qed
\end{pf}

\subsection{Identification of faulty interactions}

\begin{figure}
    \caption{$2$-CCA, $(\overline{1},2)$-CLA, and test outcomes in Case 1 and Case 2.}
    \label{fig:ccaclaresults}
    \begin{minipage}[t]{0.4\textwidth}
        \subcaption{Test outcomes of 2-CCA.}
        \centering
        \input{2ccaWithResultsCase1.tex}
    \end{minipage}
    \hfill
    \begin{minipage}[t]{0.4\textwidth}
        \subcaption{Test outcomes of $(\overline{1},2)$-CLA.}
        \centering
        \input{12claWithResultsCase1.tex}
    \end{minipage}
\end{figure}

\begin{figure}
    \caption{$(\overline{1},2)$-CDA and test outcomes in Case 1 and Case 2.}
    \label{fig:cdaresults}
    \centering
    \input{12cdaWithResultsCase1.tex}
\end{figure}

Using a CDA as a test suite, faulty interactions are identified as follows. 
The execution of a test suite yields a test outcome which is a set of failed test cases and a set of passing test cases. 
Every interaction of the target strength is identified as faulty iff it appears in at least one of the failed test cases but none of the passing ones. 
For a $(\overline{d}, t)$-CDA or a $(\overline{d}, \overline{t})$-CDA, the target strength is $t$ or $0, 1, \dots, t$, respectively. 
Let $\mathcal{T}$ be the set of the faulty interactions. 
In other words, for a CDA $A$ and an interaction $T$ of the target strength, 
$T$ is identified as faulty iff $\rho_A(T) \subseteq \rho_A(\mathcal{T})$. 
Note that we do not know $\mathcal{T}$ in advance but know the set of failed test cases $\rho_A(\mathcal{T})$.

Suppose that a $(d, t)$-CDA $A$ is used as a test suite assuming 
that the number of faulty interactions is at most $d$ 
and that they are all $t$-way. 

First let us consider the case where the assumptions are indeed true. 
For an interaction $T \in \mathcal{VI}_t$, if $\mathcal{T} \not \succ T$, then 
$T \in \mathcal{T} \Leftrightarrow \rho_A(T) \subseteq \rho_A(\mathcal{T})$ holds; i.e.,  
$T$ is faulty iff test cases in which $T$ appears all failed.
Hence $T$ is accurately determined to be faulty or not faulty, except when 
 $\mathcal{T} \succ T$, in which case $T$ is identified as faulty. 

Next consider the case where the assumption on the number of faulty interactions is false. 
That is, there are more than $d$ faulty interactions.
In this case, interactions $T$ can be falsely determined to be faulty even if $\mathcal{T} \not \succ T$. 
However, all faulty interactions are correctly identified as faulty, 
because all valid $t$-way interactions appear in $A$ (Theorem~\ref{thr:cda2cca}).
When the assumption on the strength is false, it is not possible to identify all faulty interactions. 
This is because in that case some faulty interaction may not appear in any test case, unless an exhaustive test suite is used.

The situation is similar when $A$ is a $(\overline{d}, \overline{t})$-CDA. 
In this case, the assumptions are: 
$|\mathcal{T}| \leq d$ 
and 
the strength of the faulty interactions is at most $t$. 

When these assumptions are true, $T (\in \overline{\mathcal{VI}_t})$ is accurately determined to be faulty or not faulty unless $\mathcal{T} \succ T$ or $\mathcal{T} \cup \{T\}$ is not independent,
since $T \in \mathcal{T} \Leftrightarrow \rho_A(T) \subseteq \rho_A(\mathcal{T})$.

But even when $\mathcal{T} \cup \{T\}$ is not independent, accurate identification is still possible 
if $\mathcal{T}$ contains neither proper subsets nor proper supersets of $T$ and $\mathcal{T}\not \succ T$.  
In that case, if we let $\mathcal{T}_{\min} =\{T' \in \mathcal{T} : T'' \not\subset T' \mathrm{~for~all~} T''\in \mathcal{T}\}$ (i.e., $\mathcal{T}_{\min} (\subseteq \mathcal{T})$ is the set of minimal interactions in $\mathcal{T}$), then 
$\mathcal{T}_{\min} \not \succ T$ and $\mathcal{T}_{\min} \cup \{T\}$ becomes independent, and thus
$T \in \mathcal{T}_{\min} \Leftrightarrow \rho_A(T) \in \rho_A(\mathcal{T}_{\min})$. 
Also $T \in \mathcal{T} \Leftrightarrow T \in \mathcal{T}_{\min}$ and $\rho_A(\mathcal{T}) = \rho_A(\mathcal{T}_{\min})$. 
Consequently $T \in \mathcal{T} \Leftrightarrow \rho_A(T) \subseteq \rho_A(\mathcal{T})$. 

In sum, faulty interactions are all identified as faulty 
and a non-faulty interaction may be identified as faulty 
only if the set of the faulty interactions masks the non-faulty interaction or contains 
its proper subsets or supersets. 

When $|\mathcal{T}| > d$, 
non-faulty interactions might be falsely identified as faulty; but 
all faulty interactions are correctly identified as faulty.
The faulty interactions can be correctly identified because they only appear in the failed test cases. 


% ...

When the assumption on the strength is not false, it is not possible to identify all faulty interactions.


\subsubsection{Examples}

Here using the running example, we illustrate how CCA, CLA and CDA arrays are used to detect and locate faulty interactions. 
We consider the cases $d=1$ and  $t=2$. 
Suppose that the $2$-way CCA, the $(1,2)$-CLA, and the $(1,2)$-CDA shown in Figures~\ref{fig:2cca}, \ref{fig:12cla}, and Figure~\ref{fig:12cda} are used as test suites. 
In fact the CLA and the CDA are a $(\overline{1}, 2)$-CLA and a $(\overline{1}, 2)$-CDA. 
Figure~\ref{fig:ccaclaresults} and Figure~\ref{fig:cdaresults} summarize the results of test cases when executed in the two cases below. 

\begin{enumerate}[\textit{Case} 1]
    \item The only faulty interaction is $T_\alpha=\{(F_1,0),(F_2,0)\}$. 
    \item There are two faulty interactions $T_\beta=\{(F_1,0),(F_3,0)\}$ and $T_\gamma=\{(F_1,0),(F_4,1)\}$. 
        
\end{enumerate}

\paragraph{CCA}

In \textit{Case}~1, within the test cases in the 2-CCA~(Figure~\ref{fig:2cca}), only the test cases $\boldsymbol{\sigma}_1$ and $\boldsymbol{\sigma}_2$ fail. 
The two-way interactions that appear only in those failed test cases are as follows
(the faulty interaction is indicated by underline.)
\begin{equation*}
    \begin{aligned}
        &\underline{\{(F_1,0),(F_2,0)\}}\qquad\{(F_1,0),(F_3,0)\}\qquad\{(F_1,0),(F_4,0)\}\\
        &\{(F_2,0),(F_4,0)\}\qquad\{(F_3,0),(F_4,0)\}\qquad\{(F_1,0),(F_4,3)\}\\
    \end{aligned}
\end{equation*}
In \textit{Case}~2, the failed test cases are $\sigma_1$, $\sigma_2$, and $\sigma_3$; thus the candidates for faulty interactions are:
\begin{equation*}
    \begin{aligned}
    &\{(F_1,0),(F_2,0)\} \qquad \underline{\{(F_1,0),(F_3,0)\}} \qquad \{(F_1,0),(F_4,0)\}\\
    &\{(F_2,0),(F_4,0)\} \qquad \{(F_3,0),(F_4,0)\} \qquad \{(F_1,0),(F_4,3)\}\\
    &\{(F_1,0),(F_3,1)\} \qquad \underline{\{(F_1,0),(F_4,1)\}} \qquad \{(F_2,1),(F_4,1)\}\\
    &\{(F_3,1),(F_4,1)\} \\
\end{aligned}
\end{equation*}
For both cases it is impossible to further reduce the candidates of faulty interactions.



\paragraph{CLA}
Suppose that the $(1,2)$-CLA~($\overline{1},2$-CLA) shown in Figure~\ref{fig:12cla} is used. 
In \textit{Case} 1, the test cases $\boldsymbol{\sigma}_1$, $\boldsymbol{\sigma}_2$ and $\boldsymbol{\sigma}_3$ 
fail and all the other test cases pass.
The interactions that appear only in the failed test cases are as follows. 
\begin{equation*}
    \begin{aligned}
        & \underline{\{(F_1,0),(F_2,0)\}}\qquad\{(F_1,0),(F_3,0)\}\qquad\{(F_3,0),(F_4,0)\}\\ 
        & \{(F_1,0),(F_4,3)\}\qquad\{(F_1,0),(F_4,1)\} \\
    \end{aligned}
\end{equation*}
The core idea of CLAs is that it allows a test outcome to be uniquely associated with a set of faulty interactions, 
which is mathematically represented as $\mathcal{T}_1 = \mathcal{T}_2 \Leftrightarrow \rho_A(\mathcal{T}_1) = \rho_A(\mathcal{T}_2)$. 
In this case, $\rho_A(\mathcal{T}) = \rho_A(\{T_\alpha\})= \{\sigma_1, \sigma_2, \sigma_3\}$ holds only for $\mathcal{T} = \{ \{(F_1,0),(F_2,0)\} \}$,
provided that  $\mathcal{T} \subseteq \mathcal{VI}_2$ and $|\mathcal{T}| \leq 1$.
%
Thus, we correctly locate the faulty interaction. 

Now consider \textit{Case}~2. 
The failed test cases are the same as in \textit{Case}~1, i.e., $\boldsymbol{\sigma}_1$, $\boldsymbol{\sigma}_2$ and $\boldsymbol{\sigma}_3$. 
Hence the conclusion that $T_\alpha$ is the only faulty interaction is also the same. 
This incorrect result is caused by that the number of faulty interactions does not coincide with the assumption (namely $d = 1$). 
In general, if faulty interactions exceed the number assumed, CLAs may identify non-faulty interactions as faulty but also identify faulty interactions as non-faulty. 

%

\paragraph{CDA}

Suppose that the  $(1,2)$-CDA~($(\overline{1},2)$-CDA) shown in Figure~\ref{fig:12cda} is used to locate faulty interactions. 
For \textit{Case}~1, the failed test cases are $\boldsymbol{\sigma}_1$, $\boldsymbol{\sigma}_2$, $\boldsymbol{\sigma}_3$, $\boldsymbol{\sigma}_4$ and $\boldsymbol{\sigma}_5$. 
The interactions occurring only in the failed test cases are all identified as faulty. 
In this case these interactions are:
\begin{equation*}
    \begin{aligned}
        &\underline{\{(F_1,0),(F_2,0)\}}\qquad\{(F_1,0),(F_3,0)\}\qquad\{(F_1,0),(F_4,3)\}
    \end{aligned}
\end{equation*}
$T_\alpha$ is correctly identified as faulty, whereas $\{(F_1,0),(F_3,0)\}$ and $\{(F_1,0),(F_4,3)\}$ are incorrectly identified as faulty. 
Since $\{T_a\}$ masks $\{(F_1,0),(F_3,0)\}$ and $\{(F_1,0),(F_4,3)\}$ ($\{T_\alpha\} \succ \{(F_1,0),(F_3,0)\}$ and $\{T_\alpha\} \succ \{(F_1,0),(F_4,3)\}$), 
it is inherently impossible to determine that $\{(F_1,0),(F_3,0)\}$ and $\{(F_1,0),(F_4,3)\}$ are not faulty when $T_\alpha$ is faulty. 
However, it should be noted that if we relied on the assumption that the number of faulty interactions is $d = 1$, just as in the case of the CLA above, 
we could correctly identify only $T_\alpha$ as faulty. 
In fact, we will show later that any CDA is a CLA. 

For \textit{Case}~2, the failed test cases are $\boldsymbol{\sigma}_1$, $\boldsymbol{\sigma}_2$, $\boldsymbol{\sigma}_3$,
$\boldsymbol{\sigma}_4$,
and  $\boldsymbol{\sigma}_8$. 
The interactions that are identified as faulty are:
\begin{equation*}
    \begin{aligned}
        &\underline{\{(F_1,0),(F_3,0)\}}\qquad\underline{\{(F_1,0),(F_4,1)\}}\qquad\{(F_1,0),(F_4,3)\}\\
    \end{aligned}
\end{equation*}
Although the last interaction is in fact not faulty, all the faulty ones are correctly identified. 
In general, when using a CDA, non-faulty interactions are never wrongly identified as faulty even if the number 
of faulty interactions exceeds the assumed number $d$. 

\subsection{Properties of CDAs}

In the rest of the section we provide some theorems on the properties of CDAs. 
\begin{theorem}
    $\mathcal{R}$, the exhaustive test suite, is a $(d,t)$-, $(d,\overline{t})$-, $(\overline{d},t)$- and $(\overline{d},\overline{t})$-CDA for any $d$ and $t$.
\end{theorem}

\begin{pf}
Let $T$ be a valid interaction and $\mathcal{T}$ be a set of valid interactions. 
Below we will show $\mathcal{T} \not\succ T \Rightarrow (T \in \mathcal{T} \Leftrightarrow \rho_{\mathcal{R}}(T) \subseteq \rho_{\mathcal{R}}(\mathcal{T}))$.
If  $\mathcal{T} \not\succ T$ and $T \not\in \mathcal{T}$, then there is some $\sigma \in \mathcal{R}$ such that $T \subseteq \sigma$ and $\forall T'\in \mathcal{T}: T' \not \subseteq \sigma$, 
in which case $\sigma \in \rho_{\mathcal{R}}(T) - \rho_{\mathcal{R}}(\mathcal{T})$. 
That is, $\mathcal{T} \not\succ T \Rightarrow (T \not \in \mathcal{T} \Rightarrow \rho_{\mathcal{R}}(T) \not \subseteq \rho_{\mathcal{R}}(\mathcal{T}))$. 
In addition $T \in \mathcal{T} \Rightarrow \rho_{\mathcal{R}}(T) \subseteq \rho_{\mathcal{R}}(\mathcal{T})$ trivially holds.  
As a result, the theorem follows. \qed
\end{pf}


\begin{theorem}
    \label{theo:equiv3}
    A $(d,t)$-CDA is also a $(d,t)$-CLA; a $(\overline{d},t)$-CDA is also a $(\overline{d},t)$-CLA; A $(d,\overline{t})$-CDA is also a $(d,\overline{t})$-CLA and a $(\overline{d},\overline{t})$-CDA is also a $(\overline{d},\overline{t})$-CLA. 
\end{theorem}

\begin{pf}
    Let $A$ be a $(d,t)$-CDA. 
Let $\mathcal{T}_1$ and $\mathcal{T}_2$ be different sets of $t$-way interactions of size $d$ and mutually distinguishable.
$\rho_{\mathcal{R}}(\mathcal{T}_1)\neq\rho_{\mathcal{R}}(\mathcal{T}_2)$ by the definition of distinguishability. 
Without loss of generality, we assume that $\mathcal{T}_1 \not \subset \mathcal{T}_2$. 
Denote the test case that exists in $\rho_{\mathcal{R}}(\mathcal{T}_1)$ but not in $\rho_{\mathcal{R}}(\mathcal{T}_2)$ as $\tc_e$. There exists at least one valid interaction $T$ in $\mathcal{T}_1$ that is covered by $\tc_e$. 
$\mathcal{T}_2$ does not mask $T$ and $T \not \in \mathcal{T}_2$ because $\tc_e\in\rho_{\mathcal{R}}(T)\land\tc_e\not\in\rho_{\mathcal{R}}(\mathcal{T}_2)$. 
Since $A$ is a $(d,t)$-CDA, $\mathcal{T}_2\not\succ T$, 
and $T \not \in \mathcal{T}_2$, we have $\rho_A(T) \not\subseteq \rho_A(\mathcal{T}_2)$.
Hence there exists a row $\tc_e'$ in $A$ such that $\tc_e'$ covers $T$ but does not cover any interactions in $\mathcal{T}_2$, that is, $\tc_e'\in\rho_A(\mathcal{T}_1)\land\tc_e'\not\in\rho_A(\mathcal{T}_2)$ holds. Thus $\rho_A(\mathcal{T}_1)\neq\rho_A(\mathcal{T}_2)$ holds; hence $A$ is a $(d, t)$-CLA.
The same argument holds when $|\mathcal{T}_1|$ and
$|\mathcal{T}_2|$ are not exactly $d$ but at most $d$. Therefore 
it follows that a $(\overline{d},t)$-CDA is a $(\overline{d},t)$-CLA.

Next let $A$ be a $(d, \overline{t})$-CDA or a $(\overline{d}, \overline{t})$-CDA. 
Let $\mathcal{T}_1$ and $\mathcal{T}_2$ be different sets consisting of exactly $d$ interactions of strength at most $t$
or at most $d$ interactions of strength at most $t$, respectively. 
Then the same argument for the case of $(d, t)$-CDAs and  $(\overline{d}, t)$-CDAs holds. 
As a result, the theorem follows. \qed
\end{pf}

\begin{theorem}
    \label{theorem:cca2cda}
A $(t+d)$-CCA is a $(\overline{d}, \overline{t})$-CDA.
\end{theorem}

\begin{pf}

Suppose that A is a $(t+ d)$-CCA. 
The theorem holds if $\rho_A(T) \not\subseteq \rho_A(\mathcal{T})$ for any $T \in \overline{\mathcal{VI}_t}$ and $\mathcal{T}\subseteq \overline{\mathcal{VI}_t}$ 
such that $0 \leq |\mathcal{T}| \leq d$, $T\not \in \mathcal{T}$, $\mathcal{T}\not\succ T$, and $\{T\}\cup \mathcal{T}$ is independent. 
We show this by constructing a valid interaction $\hat T$ of strength $\leq d + t$ that covers $T$ but cannot appear with any interaction in $\mathcal{T}$ in the same row. 
If such $\hat T$ exists, some row of $A$ contains it because $A$ is a $(t+d)$-CCA. This row is in $\rho_A(T)$ but not in $\rho_A(\mathcal{T})$; thus $\rho_A(T) \not \subseteq \rho_A(\mathcal{T})$.

Since $\mathcal{T}\not\succ T$, there must be a valid test case $\sigma$ that covers $T$ but does not cover any $T'\in \mathcal{T}$. Let $\sigma = \langle s_1, s_2, \dots, s_k\rangle$. We regard $\sigma$ as $k$-way interaction $\{ (F_1, s_1), (F_2, s_2),\dots, (F_k, s_k)\}$. $\hat T$ is constructed by starting from $\hat T = T$ and gradually expanding it by applying the following process for all $T'\in \mathcal{T}$: Select any $(F_i, v) \in T'$ such that $s_i \neq v$. This can be done because $T'$ is not covered by $\sigma$ (and thus $\curlywedge \not\in \mathcal{T}$). Add $(F_i, s_i)$ to $\hat T$. Finally $\hat T$ becomes the desired interaction. \qed
\end{pf}

%% file: 11cda.tex
\begin{tabular}{c|cccc}
\hline 
 & $F_1$ & $F_2$ & $F_3$ & $F_4$\tabularnewline
\hline 
$\boldsymbol\sigma_1$ & 0 & 0 & 0 & 0\tabularnewline
$\boldsymbol\sigma_2$ & 0 & 1 & 2 & 1\tabularnewline
$\boldsymbol\sigma_3$ & 1 & 0 & 0 & 3\tabularnewline
$\boldsymbol\sigma_4$ & 1 & 0 & 1 & 1\tabularnewline
$\boldsymbol\sigma_5$ & 1 & 1 & 1 & 2\tabularnewline
$\boldsymbol\sigma_6$ & 2 & 0 & 0 & 3\tabularnewline
$\boldsymbol\sigma_7$ & 2 & 0 & 2 & 2\tabularnewline
$\boldsymbol\sigma_8$ & 2 & 1 & 1 & 0\tabularnewline
\hline 
\end{tabular}

%% file: 21cda.tex
\begin{tabular}{c|cccc}
\hline 
 & $F_1$ & $F_2$ & $F_3$ & $F_4$\tabularnewline
\hline 
$\boldsymbol\sigma_1$ & 0 & 0 & 0 & 0\tabularnewline
$\boldsymbol\sigma_2$ & 0 & 0 & 0 & 3\tabularnewline
$\boldsymbol\sigma_3$ & 0 & 0 & 2 & 1\tabularnewline
$\boldsymbol\sigma_4$ & 0 & 1 & 1 & 2\tabularnewline
$\boldsymbol\sigma_5$ & 0 & 1 & 2 & 0\tabularnewline
$\boldsymbol\sigma_6$ & 1 & 0 & 1 & 1\tabularnewline
$\boldsymbol\sigma_7$ & 1 & 0 & 0 & 1\tabularnewline
$\boldsymbol\sigma_8$ & 1 & 0 & 0 & 2\tabularnewline
$\boldsymbol\sigma_9$ & 1 & 0 & 0 & 3\tabularnewline
$\boldsymbol\sigma_{10}$ & 1 & 1 & 1 & 0\tabularnewline
$\boldsymbol\sigma_{11}$ & 1 & 1 & 2 & 1\tabularnewline
$\boldsymbol\sigma_{12}$ & 2 & 0 & 0 & 3\tabularnewline
$\boldsymbol\sigma_{13}$ & 2 & 0 & 1 & 2\tabularnewline
$\boldsymbol\sigma_{14}$ & 2 & 0 & 2 & 0\tabularnewline
$\boldsymbol\sigma_{15}$ & 2 & 1 & 1 & 1\tabularnewline
$\boldsymbol\sigma_{16}$ & 2 & 1 & 2 & 2\tabularnewline
\hline 
\end{tabular}

%% file: 12cda.tex
\begin{tabular}{c|cccc}
\hline 
 & $F_1$ & $F_2$ & $F_3$ & $F_4$\tabularnewline
\hline 
$\boldsymbol\sigma_1$ & 0 & 0 & 0 & 0\tabularnewline
$\boldsymbol\sigma_2$ & 0 & 0 & 0 & 1\tabularnewline
$\boldsymbol\sigma_3$ & 0 & 0 & 0 & 3\tabularnewline
$\boldsymbol\sigma_4$ & 0 & 0 & 1 & 1\tabularnewline
$\boldsymbol\sigma_5$ & 0 & 0 & 2 & 2\tabularnewline
$\boldsymbol\sigma_6$ & 0 & 1 & 1 & 2\tabularnewline
$\boldsymbol\sigma_7$ & 0 & 1 & 2 & 0\tabularnewline
$\boldsymbol\sigma_8$ & 0 & 1 & 2 & 1\tabularnewline
$\boldsymbol\sigma_9$ & 1 & 0 & 0 & 2\tabularnewline
$\boldsymbol\sigma_{10}$ & 1 & 0 & 0 & 3\tabularnewline
$\boldsymbol\sigma_{11}$ & 1 & 0 & 1 & 0\tabularnewline
$\boldsymbol\sigma_{12}$ & 1 & 0 & 2 & 1\tabularnewline
$\boldsymbol\sigma_{13}$ & 1 & 1 & 1 & 1\tabularnewline
$\boldsymbol\sigma_{14}$ & 1 & 1 & 2 & 0\tabularnewline
$\boldsymbol\sigma_{15}$ & 1 & 1 & 2 & 2\tabularnewline
$\boldsymbol\sigma_{16}$ & 2 & 0 & 0 & 0\tabularnewline
$\boldsymbol\sigma_{17}$ & 2 & 0 & 0 & 1\tabularnewline
$\boldsymbol\sigma_{18}$ & 2 & 0 & 0 & 2\tabularnewline
$\boldsymbol\sigma_{19}$ & 2 & 0 & 0 & 3\tabularnewline
$\boldsymbol\sigma_{20}$ & 2 & 0 & 1 & 2\tabularnewline
$\boldsymbol\sigma_{21}$ & 2 & 0 & 2 & 0\tabularnewline
$\boldsymbol\sigma_{22}$ & 2 & 1 & 1 & 0\tabularnewline
$\boldsymbol\sigma_{23}$ & 2 & 1 & 1 & 1\tabularnewline
$\boldsymbol\sigma_{24}$ & 2 & 1 & 2 & 2\tabularnewline
\hline 
\end{tabular}

%% file: 2ccaWithResultsCase1.tex
\begin{tabular}{c|ccccccc}
\cline{1-5} \cline{2-5} \cline{3-5} \cline{4-5} \cline{5-5} \cline{7-8} \cline{8-8} 
 & $F_1$ & $F_2$ & $F_3$ & $F_4$ &  & Case1 & Case2\tabularnewline
\cline{1-5} \cline{2-5} \cline{3-5} \cline{4-5} \cline{5-5} \cline{7-8} \cline{8-8}
$\boldsymbol\sigma_1$ & 0 & 0 & 0 & 0 &  & \textbf{Fail} & \textbf{Fail} \tabularnewline
$\boldsymbol\sigma_2$ & 0 & 0 & 0 & 3 &  & \textbf{Fail} & \textbf{Fail}\tabularnewline
$\boldsymbol\sigma_3$ & 0 & 1 & 1 & 1 &  & Pass & \textbf{Fail}\tabularnewline
$\boldsymbol\sigma_4$ & 0 & 1 & 2 & 2 &  & Pass & Pass\tabularnewline
$\boldsymbol\sigma_5$ & 1 & 0 & 0 & 2 &  & Pass & Pass\tabularnewline
$\boldsymbol\sigma_6$ & 1 & 0 & 0 & 3 &  & Pass & Pass\tabularnewline
$\boldsymbol\sigma_7$ & 1 & 0 & 2 & 1 &  & Pass & Pass\tabularnewline
$\boldsymbol\sigma_8$ & 1 & 1 & 1 & 0 &  & Pass & Pass\tabularnewline
$\boldsymbol\sigma_9$ & 2 & 0 & 0 & 1 &  & Pass & Pass\tabularnewline
$\boldsymbol\sigma_{10}$ & 2 & 0 & 0 & 3 &  & Pass & Pass\tabularnewline
$\boldsymbol\sigma_{11}$ & 2 & 0 & 1 & 2 &  & Pass & Pass\tabularnewline
$\boldsymbol\sigma_{12}$ & 2 & 1 & 2 & 0 &  & Pass & Pass\tabularnewline
\cline{1-5} \cline{2-5} \cline{3-5} \cline{4-5} \cline{5-5} \cline{7-8} \cline{8-8}
\end{tabular}

%% file: 12claWithResultsCase1.tex
\begin{tabular}{c|ccccccc}
\cline{1-5} \cline{2-5} \cline{3-5} \cline{4-5} \cline{5-5} \cline{7-8}  \cline{8-8} 
 & $F_1$ & $F_2$ & $F_3$ & $F_4$ &  & Case1 & Case2\tabularnewline
\cline{1-5} \cline{2-5} \cline{3-5} \cline{4-5} \cline{5-5} \cline{7-8}  \cline{8-8} 
$\boldsymbol\sigma_1$ & 0 & 0 & 0 & 0 &  & \textbf{Fail} & \textbf{Fail}\tabularnewline
$\boldsymbol\sigma_2$ & 0 & 0 & 0 & 3 &  & \textbf{Fail} & \textbf{Fail}\tabularnewline
$\boldsymbol\sigma_3$ & 0 & 0 & 1 & 1 &  & \textbf{Fail} & \textbf{Fail}\tabularnewline
$\boldsymbol\sigma_4$ & 0 & 1 & 1 & 2 &  & Pass & Pass\tabularnewline
$\boldsymbol\sigma_5$ & 0 & 1 & 2 & 0 &  & Pass & Pass\tabularnewline
$\boldsymbol\sigma_6$ & 1 & 0 & 0 & 2 &  & Pass & Pass\tabularnewline
$\boldsymbol\sigma_7$ & 1 & 0 & 0 & 3 &  & Pass & Pass\tabularnewline
$\boldsymbol\sigma_8$ & 1 & 0 & 1 & 2 &  & Pass & Pass\tabularnewline
$\boldsymbol\sigma_9$ & 1 & 1 & 1 & 1 &  & Pass & Pass\tabularnewline
$\boldsymbol\sigma_{10}$ & 1 & 1 & 2 & 0 &  & Pass & Pass\tabularnewline
$\boldsymbol\sigma_{11}$ & 1 & 1 & 2 & 2 &  & Pass & Pass\tabularnewline
$\boldsymbol\sigma_{12}$ & 2 & 0 & 0 & 1 &  & Pass & Pass\tabularnewline
$\boldsymbol\sigma_{13}$ & 2 & 0 & 0 & 3 &  & Pass & Pass\tabularnewline
$\boldsymbol\sigma_{14}$ & 2 & 0 & 2 & 0 &  & Pass & Pass\tabularnewline
$\boldsymbol\sigma_{15}$ & 2 & 1 & 1 & 0 &  & Pass & Pass\tabularnewline
$\boldsymbol\sigma_{16}$ & 2 & 1 & 1 & 2 &  & Pass & Pass\tabularnewline
$\boldsymbol\sigma_{17}$ & 2 & 1 & 2 & 1 &  & Pass & Pass\tabularnewline
\cline{1-5} \cline{2-5} \cline{3-5} \cline{4-5} \cline{5-5} \cline{7-8}  \cline{8-8} 
\end{tabular}

%% file: 12cdaWithResultsCase1.tex
\begin{tabular}{c|ccccccc}
\cline{1-5} \cline{2-5} \cline{3-5} \cline{4-5} \cline{5-5} \cline{7-8} \cline{8-8} 
 & $F_1$ & $F_2$ & $F_3$ & $F_4$ &  & Case1 & Case2\tabularnewline
\cline{1-5} \cline{2-5} \cline{3-5} \cline{4-5} \cline{5-5} \cline{7-8} \cline{8-8} 
$\boldsymbol\sigma_1$ & 0 & 0 & 0 & 0 &  & \textbf{Fail} & \textbf{Fail}\tabularnewline
$\boldsymbol\sigma_2$ & 0 & 0 & 0 & 1 &  & \textbf{Fail} & \textbf{Fail}\tabularnewline
$\boldsymbol\sigma_3$ & 0 & 0 & 0 & 3 &  & \textbf{Fail} & \textbf{Fail}\tabularnewline
$\boldsymbol\sigma_4$ & 0 & 0 & 1 & 1 &  & \textbf{Fail} & \textbf{Fail}\tabularnewline
$\boldsymbol\sigma_5$ & 0 & 0 & 2 & 2 &  & \textbf{Fail} & Pass\tabularnewline
$\boldsymbol\sigma_6$ & 0 & 1 & 1 & 2 &  & Pass & Pass\tabularnewline
$\boldsymbol\sigma_7$ & 0 & 1 & 2 & 0 &  & Pass & Pass\tabularnewline
$\boldsymbol\sigma_8$ & 0 & 1 & 2 & 1 &  & Pass & \textbf{Fail}\tabularnewline
$\boldsymbol\sigma_9$ & 1 & 0 & 0 & 2 &  & Pass & Pass\tabularnewline
$\boldsymbol\sigma_{10}$ & 1 & 0 & 0 & 3 &  & Pass & Pass\tabularnewline
$\boldsymbol\sigma_{11}$ & 1 & 0 & 1 & 0 &  & Pass & Pass\tabularnewline
$\boldsymbol\sigma_{12}$ & 1 & 0 & 2 & 1 &  & Pass & Pass\tabularnewline
$\boldsymbol\sigma_{13}$ & 1 & 1 & 1 & 1 &  & Pass & Pass\tabularnewline
$\boldsymbol\sigma_{14}$ & 1 & 1 & 2 & 0 &  & Pass & Pass\tabularnewline
$\boldsymbol\sigma_{15}$ & 1 & 1 & 2 & 2 &  & Pass & Pass\tabularnewline
$\boldsymbol\sigma_{16}$ & 2 & 0 & 0 & 0 &  & Pass & Pass\tabularnewline
$\boldsymbol\sigma_{17}$ & 2 & 0 & 0 & 1 &  & Pass & Pass\tabularnewline
$\boldsymbol\sigma_{18}$ & 2 & 0 & 0 & 2 &  & Pass & Pass\tabularnewline
$\boldsymbol\sigma_{19}$ & 2 & 0 & 0 & 3 &  & Pass & Pass\tabularnewline
$\boldsymbol\sigma_{20}$ & 2 & 0 & 1 & 2 &  & Pass & Pass\tabularnewline
$\boldsymbol\sigma_{21}$ & 2 & 0 & 2 & 0 &  & Pass & Pass\tabularnewline
$\boldsymbol\sigma_{22}$ & 2 & 1 & 1 & 0 &  & Pass & Pass\tabularnewline
$\boldsymbol\sigma_{23}$ & 2 & 1 & 1 & 1 &  & Pass & Pass\tabularnewline
$\boldsymbol\sigma_{24}$ & 2 & 1 & 2 & 2 &  & Pass & Pass\tabularnewline
\cline{1-5} \cline{2-5} \cline{3-5} \cline{4-5} \cline{5-5} \cline{7-8} \cline{8-8} 
\end{tabular}

%% file: section4.tex
\section{Generation Algorithms}\label{sec:algo}

In this section, we present two algorithms for generating CDAs: the satisfiability-based algorithm and the two-step heuristic algorithm. 
In this section we limit ourselves to $(d, t)$-CDAs because $(d, t)$-CDAs are $(\overline{d}, t)$-CDAs except in extreme cases~(Theorem~\ref{theo:equiv3dbar}). Also it is straightforward to adjust the algorithms to $(d, \overline{t})$-CDAs and $(\overline{d}, \overline{t})$-CDAs.  

\subsection{The satisfiability-based algorithm}

The first algorithm leverages a satisfiability solver. 
We reduce the problem of generating a CDA of a given size to the satisfiability problem of a logical~(i.e., Boolean-valued) expression. 
A logical expression is satisfiable iff it evaluates to true for some \emph{valuation}, i.e., assignment of values to the variables. 
The algorithm first estimates the upper bound on the minimum size of a CDA and uses it as the initial size of a CDA.  
Then it creates a logical expression that is satisfiable iff a CDA of the initial size exists. 
The logical expression is in turn evaluated by a satisfiability solver. 
We design the logical expression so that the valuation that satisfies it directly represents a CDA. 
Satisfiability solvers can produce such a satisfying valuation when the expression is satisfiable; 
hence a CDA can be obtained from the output of the solver. 
Repeating the process while decreasing the CDA size, the algorithm can obtain the smallest CDA. 

\subsubsection{The logic expression}

To represent an array with a collection of variables, 
we adopt the \textit{na\"ive matrix model} which is used by Hnich~et~al.~\cite{hnich_constraint2006} in their study to find CAs.
In this model, an $N\times k$ array is represented as an $N \times k$ matrix of integer variables as follows.
\begin{equation*}
  A = \left(
  \begin{array}{ccc}
  p_1^1 & \cdots & p_k^1\\
  \vdots & \ddots & \vdots\\
  p_1^N & \cdots & p_k^N
  \end{array}
  \right)
\end{equation*}
The variable $p_i^n$ represents the value on the parameter $F_i$ in the $n$-th test case.
The domain of $p_i^n$ is $S_i = \{0, 1, \dots, |S_i|-1\}$. 

In order for the array $A$ to become a $(d,t)$-CDA, we impose the following conditions on $A$ using logical expressions.
\begin{enumerate}
  \item The rows of $A$ represent valid test cases.  
  \item $\forall\mathcal{T}\subseteq\mathcal{VI}_t\textrm{ such that }|\mathcal{T}|=d,\forall T\in\mathcal{VI}_t:$
  $\mathcal{T}\not\succ T\Rightarrow(T\in\mathcal{T}\Leftrightarrow\rho_A(T)\subseteq\rho_A(\mathcal{T}))$
\end{enumerate}
Below we present logical expressions that represent the above two conditions.
By conjuncting all the expressions, we obtain a single logical expression to be checked for satisfiability.



\paragraph{Condition~1} In $A$, the $n$-th row is expressed as a tuple of $k$ variables $(p_{1}^{n},p_{2}^{n},\dots,p_{k}^{n})$.
As defined in Section~\ref{sec:pre}, a test case is valid iff it satisfies the constraints
and the constraints are represented by $\phi$, a Boolean-valued formula over parameters $F_1, \dots, F_k$.
We let  $\phi|_{p_{1}^{n},p_{2}^{n},\dots,p_{k}^{n}}$ denote $\phi$ with each $F_i$ being replaced with $p_i^n$.
Then, the following expression enforces $A$ to only contain valid test cases.
\begin{equation*}
  \label{eqn:validate}
  \textit{Valid} :=\bigwedge_{n=1}^{N} \phi|_{p_{1}^{n},p_{2}^{n},\dots,p_{k}^{n}}
\end{equation*}

\paragraph{Condition~2} 
It is important to note that  $\mathcal{T}\not\succ T\Rightarrow(T\in\mathcal{T}\Leftrightarrow\rho_A(T)\subseteq\rho_A(\mathcal{T}))$ is equivalent to:
\[
  (\mathcal{T}\not\succ T \land T \not \in \mathcal{T})  \Rightarrow \rho_A(T) \not \subseteq \rho_A(\mathcal{T})
\]
because 
$T\in\mathcal{T}\Rightarrow\rho_A(T)\subseteq\rho_A(\mathcal{T})$ trivially holds.
Hence we can focus on the case where $\mathcal{T} \not \succ T$ and $T \not \in \mathcal{T}$.
The right part of this formula, that is, $\rho_A(T) \not \subseteq \rho_A(\mathcal{T})$ holds 
iff there is a row in $A$ that covers $T$ but none of the interactions in $\mathcal{T}$. 
This condition is represented by a logical expression as follows:
%
\begin{equation*}
\label{eqn:locate}
\textit{Locating}(\mathcal{T}, T) := 
\bigvee_{n=1}^{N}\Big(\bigwedge_{j=1}^{t}{(p_{x_j}^n=v_{x_j})}\land \neg\Big(\bigvee_{L=1}^{d}\bigwedge_{l=1}^{t}{(p_{y_{L_l}}^n=v_{y_{L_l}})}\Big)\Big)
\end{equation*}
where $\mathcal{T} = \{\{(F_{y_{1_1}},v_{y_{1_1}}),\dots,(F_{y_{1_t}},v_{y_{1_t}})\},\dots$, $\{(F_{y_{d_1}},v_{y_{d_1}})$, $\dots$, $(F_{y_{d_t}},v_{y_{d_t}})\}\}$ and $T = \{(F_{x_1},v_{x_1}),\dots,(F_{x_t},v_{x_t})\}$.
For given $\mathcal{T}$ and $T$,  $\rho_A(T) \not \subseteq \rho_A(\mathcal{T})$ holds iff $\textit{Locating}(\mathcal{T}, T)$ is satisfiable. 

Let us define $\mathcal{U}$ as follows: 
\[
\mathcal{U} := \{ (\mathcal{T}, T) \mid \mathcal{T} \subseteq \mathcal{VI}_t, |\mathcal{T}| = d, T \in \mathcal{VI}_t,  \mathcal{T} \not \succ T, T\not \in \mathcal{T}\}  
\]
By ANDing $\textit{Locating}(\mathcal{T}, T)$ for all $(\mathcal{T}, T) \in \mathcal{U}$, 
we obtain an expression that represents the second condition. 

\paragraph{The whole expression} The whole expression that will be checked for satisfiability
is obtained by conjuncting the expressions defined above as follows:
\begin{equation*}
  \textit{existCDA} := \ \textit{Valid} \ \land 
  \bigwedge_{(\mathcal{T},T) \in \mathcal{U}}\textit{Locating}(\mathcal{T},T) 
    \end{equation*}
By checking the satisfiability of this expression, whether a $(d,t)$-CDA of size $N$ exists or not can be determined.
If it is satisfiable, then a CDA of size $N$ exists. 
In this case, the satisfying valuation for the $N\times k$ variables $p_i^n$ represents all the entries of one such CDA.
On the other hand, if the expression is unsatisfiable, then it can be concluded that no $(d,t)$-CDA of size $N$ exists.


The satisfiability of the above expression can be checked using 
Constraint Satisfaction Problem (CSP) solvers, 
Satisfiability Modulo Theories (SMT) solvers, 
or Boolean Satisfiability (SAT) solvers with a Boolean encoding of integers.

\subsubsection{Computing $\mathcal{U}$}


In order to construct the above logical expression $\textit{existCDA}$, we need to obtain $\mathcal{U}$
 (see the subscript of the $\bigwedge$ in the expression).  
Computing $\mathcal{U}$ requires $\mathcal{VI}_t$. 
We will show how to compute $\mathcal{VI}_t$ later. Here we describe how one can compute $\mathcal{U}$ when $\mathcal{VI}_t$ is available.

Now consider enumerating all $\mathcal{T}$-$T$ pairs such that
$T \in\mathcal{VI}_t$, $\mathcal{T} \subseteq \mathcal{VI}_t$, $|\mathcal{T}|=d$,
$T\not\in\mathcal{T}$, and $\mathcal{T}\not\succ T$. 
The problem here is how to decide whether or not
 $\mathcal{T} (\subseteq \mathcal{VI}_t)$
masks $T (\in \mathcal{VI}_t)$ when $\mathcal{T} $ and $T \not\in \mathcal{T}$ are given.
This too is possible by making use of satisfiability solving. 
We let integer variables $p_1, p_2, \dots, p_k$ to symbolically represent
a test case $\tc$; that is,
\[
\tc = (p_1, p_2, \dots, p_k)
\]
The domain of $p_i$ is $\{0, 1, \dots, |S_i|-1\}$.
Note that $S_i$ is the domain of parameter $F_i$.

By the definition of masking, given such a $\mathcal{T}$-$T$ pair,
 $\mathcal{T}$ does not mask $T$ iff the following condition holds:
 \[
 \exists
 	\tc\in\mathcal{R}: T\subseteq\tc \land \neg (\exists T'\in\mathcal{T}: T'\subseteq\tc)
 \]
In words, the condition holds if
there is a valid test case that covers the interaction $T$ but does not cover
any interactions in the interaction set $\mathcal{T}$.
Hence, given $\mathcal{T}$ and $T$, $\mathcal{T}\not\succ T$ holds iff the following
expression evaluates to true.
\begin{equation*}
\label{eqn:unmask}
\begin{aligned}
&\textit{checkUnMasking}(\mathcal{T}, T):=\\
&\bigwedge_{j=1}^{t}{(p_{x_j}=v_{x_j})}\land \neg\Big(\bigvee_{L=1}^{d}\bigwedge_{l=1}^{t}{(p_{y_{L_l}}=v_{y_{L_l}})}\Big)\land \phi|_{p_1,p_2,\dots,p_k}
\end{aligned}
\end{equation*}
where $\mathcal{T} = \{\{(F_{y_{1_1}},v_{y_{1_1}}),\dots,(F_{y_{1_t}},v_{y_{1_t}})\},\dots$, $\{(F_{y_{d_1}},v_{y_{d_1}})$, $\dots$, $(F_{y_{d_t}},v_{y_{d_t}})\}\}$ and $T = \{(F_{x_1},v_{x_1}),\dots,(F_{x_t},v_{x_t})\}$.
$\mathcal{U}$ is obtained by, for every $\mathcal{T}$-$T$ pair, 
checking the satisfiability of $\textit{checkUnMasking}(\mathcal{T}, T)$ and keeping the pair in $\mathcal{U}$ if the expression is satisfiable. 

\subsubsection{The algorithm}

\begin{algorithm}[tb]
\caption{The satisfiability-based algorithm\label{alg:process}}
 \DontPrintSemicolon
	\KwIn{SUT $\mathcal{M}=\langle\mathcal{F},\mathcal{S},\phi\rangle$; integers $d$, $t$}
	\KwOut{\mbox{($d$,$t$)}-CDA  $A$}
  \nonl// construct a $(d+t)$-CCA for the input SUT\;
  $S\gets$\textit{generateCCA}($\mathcal{M}$, $d+t$)\label{line:cca}\;
  \nonl// get all valid $t$-way interactions from the $(d+t)$-CCA\;
  ${\cal VI}_t \gets \textit{getAllInteractions}(S, t)$\;
  \nonl// get all non-masking pairs $\mathcal{U}$ of interaction sets and interactions\;
  $\mathcal{U} \gets \mathit{getU}(\mathcal{VI}_t, d, t)$ \;
  \nonl// get the initial size for the CDA to be generated\;
  $N\gets$ The size of $S-1$ \;
  $nextA\gets S$\;
  \Do{$nextA\neq\perp$}{
      \nonl// reserve the current test suite instance\;
      $A\gets nextA$\;
      \nonl// SAT checking; the solver returns an instance if satisfiable; an emptyset, otherwise\;
      $nextA\gets$\textit{generateCDA}($\mathcal{M},d,t,N,\mathcal{U}$)\; \label{line:cda}
      \nonl// decrease the size by one\;
      $N\gets N-1$
  }
  \Return{$A$}
\end{algorithm}

The CDA generation algorithm that uses satisfiability solving is shown as Algorithm~\ref{alg:process}.
The algorithm repeatedly solves the problem of finding a $(d,t)$-CDA while varying the array size $N$.
The array size $N$ starts with a value large enough to ensure the existence of a CDA and is gradually decreased until
no existence of a CDA of size $N$ is proved.
To obtain the initial value of $N$, the algorithm creates a $(d+t)$-CCA using an off-the-shelf algorithm (line~\ref{line:cca}), where the CCA generation algorithm is represented as
function \textit{generateCCA($\mathcal{M}$,x)} which returns an $x$-CCA.
Our algorithm uses the size of the CCA minus one as the initial $N$, as any $(d+t)$-CCA is 
a $(d, t)$-CDA. 
The $(d+t)$-CCA is also used for computing $\mathcal{VI}_t$, since all valid $t$-way 
interactions appear in the CCA: The algorithm enumerates all $t$-way interactions occurring in the array, thus obtaining $\mathcal{VI}_t$. 

In the algorithm, \textit{generateCDA($\mathcal{M}$, $d$, $t$, $N$, $ \mathcal{U}$)} in line~\ref{line:cda}
represents a function that produces a $(d, t)$-CDA of size $N$
by checking the satisfiability of the expression $\textit{existCDA}$.
If the expression is satisfiable, then the SMT solver returns the satisfying valuation, 
in which case a $(d,t)$-CDA of size $N$ is obtained, since the valuation represents the $(d,t)$-CDA.
The size $N$ is decreased by one and the same process is repeated.
If the result of satisfiability check is UNSAT (unsatisfiable),  no CDA of size $N$ exists (denoted as $\perp$ in the algorithm). Then the algorithm returns the CDA of size $N+1$ and stops its execution.

One might think that binary search could work better to vary $N$ than the linear search adopted by the algorithm. In fact, this is not the case because showing unsatisfiability, that is,
the nonexistence of a CDA, usually
takes much longer time than showing satisfiability, that is, the existence of a CDA.
The linear search delays solving an unsatisfiable expression
until all possible sizes are checked, avoiding getting trapped in a long computation required for the unsatisfiable problem instance.

The size of the expression $\textit{existCDA}$ increases polynomially in $k$ when $t, d, |S_i|$, and $N$ are fixed.  
The expression can be expressed as a Boolean formula with a polynomial size increase, as $|S_i|$ is fixed. 
The Boolean satisfiability problem (SAT) is NP-complete in general and
there is no reason that SAT can be solved in polynomial-time for this particular case.  
Hence the time complexity of the algorithm is likely to be exponential.

\subsection{The two-step heuristic algorithm}

In this subsection, we propose a two-step heuristic algorithm for the generation of $(d,t)$-CDAs which aims to generate $(d,t)$-CDAs that are not optimal but fairly small in reasonable time.

Theorem~\ref{theorem:cca2cda} shows that a $(d+t)$-CCA is already a $(d,t)$-CDA. 
%
%
%
%
%
%
Based on the theorem, we devise a two-step heuristic algorithm~(Algorithm~\ref{alg:main}). 
The algorithm generates a $(d+t)$-CCA first. Then it repeatedly chooses a test case in it at random and checks whether it is removable.
Here we say that a test case is removable from an array if a new array with the test case being removed would still be a $(d, t)$-CDA. 
If the test case is removable, then it is removed from the current array. 
Otherwise, a new test case is chosen and the check is performed again. 
This process is repeated until no test case is removable anymore. 

\begin{algorithm}[tb]
\caption{The two-step heuristic generation algorithm \label{alg:main}}
 \DontPrintSemicolon
 \KwIn{SUT $\mathcal{M}=\langle\mathcal{F},\mathcal{S},\phi\rangle$; integers $d$, $t$}
	\KwOut{\mbox{($d,t$)}-CDA $A$}
  \nonl// construct a $(d+t)$-CCA for the input SUT\;
  $S \gets \textit{generateCCA}({\cal M}, d+t)$\;
  \nonl// get all t-way interactions from the $(d+t)$-CCA\;
  ${\cal VI}_t \gets \textit{getAllInteractions}(S, t)$\;
  \nonl// $Rows[T] = \rho_S(T)$ for $T \in VI_t$ \;
  $\textit{Rows}[] \gets \textit{mapInteractionToRows}({\cal VI}_t, S)$\;
  \nonl// $\textit{DiffRows}[{\cal T}][T] = \rho_S(T) - \rho_S(\mathcal{T})$ for 
  $\mathcal{T} \subseteq \mathcal{VI}_t, |\mathcal{T}|=d$\;
  $\textit{DiffRows}[][] \gets \textit{getDiffRows}({\cal VI}_t, S, d)$\;
  $A \gets S$\;
	\While{$A \not= \emptyset$}{
    $\boldsymbol{\sigma}$~$\gets$~\textit{getRandomTestcase}($S$)\;
    $S \gets S - \{\boldsymbol{\sigma}\}$\;
    $A \gets A - \{\boldsymbol{\sigma}\}$\;
    $\textit{DiffRows}'[ ][ ] \gets \textit{update}(\textit{DiffRows}[ ][ ], \boldsymbol{\sigma} )$\;
    \If{$\exists \mathcal{T}, T: \textit{DiffRows}[\mathcal{T}][T]\neq \emptyset$ 
    and $\textit{DiffRows}'[\mathcal{T}][T] = \emptyset$} {
      \nonl// the test case $\boldsymbol{\sigma}$ is unremovable\;
      $S \gets S + \{\boldsymbol{\sigma}\}$\;
    }\Else{
      \nonl// the test case $\boldsymbol{\sigma}$ is removable\;
      $\textit{DiffRows}[][] \gets \textit{DiffRows}'[][]$\;
    }
  }
  \Return{$A$}
\end{algorithm}

The algorithm works in detail as follows. 
In line~1 the algorithm generates a $(d+t)$-CCA $S$. 
At this point, $S$ is already a $(d,t)$-CDA but contains many redundant test cases.
Then the algorithm collects all valid $t$-way interactions and maps each interaction $T$ to its covering test cases $\rho_S(T)$ in $S$~(line 2).
The map obtained here, denoted by $Rows[]$, is used to compute another map, $\textit{DiffRows}[][]$,  that associates each pair of 
an interaction set $\mathcal{T}$ and a valid interaction $T$ with $\rho_S(T) - \rho_S(\mathcal{T})$. 
Note that $\rho_S(T) - \rho_S(\mathcal{T}) = \emptyset$ iff $\rho_S(T) \subseteq \rho_S(\mathcal{T})$. 
Since $S$ is a CDA, $\textit{DiffRows}[\mathcal{T}][T] = \emptyset$ if $\mathcal{T} \succ T$; $\textit{DiffRows}[\mathcal{T}][T] \not= \emptyset$ otherwise.


Then the algorithm repeatedly chooses a test case at random and checks whether it is removable or not.
To perform the check, the algorithm constructs a new interaction-to-row map $\textit{DiffRows}[][]$ that would hold after the test case was removed~(line~9). 
This can be done by simply removing $\mathbf{\sigma}$ from all $\textit{DiffRows}[\mathcal{T}][T]$.
Subsequently, the algorithm compares the two maps (line~10). 
If $\textit{DiffRows}[\mathcal{T}][T] \not = \emptyset$ but $\textit{DiffRows}'[\mathcal{T}][T] = \emptyset$, then $\rho_S(T) \subseteq \rho_S(\mathcal{T})$ and thus $S$ is no longer a CDA. 
In this case, the algorithm reserves the test case (line~12). Otherwise, it deletes the test case and accordingly updates $\textit{DiffRows}[\mathcal{T}][T]$~(line 14).
When all test cases in the CCA are checked, the algorithm will terminate, yielding the resulting $S$.


Let $s = \max_{1\leq i \leq k}{|S_i|}$. 
Outside the while loop, line~4 has the highest time complexity. 
It is $O((s^t k^t)^d s^t k^t n)$, since $|\mathcal{VI}_t| \leq s^t k^t$, $|\rho_S()| \leq n$. 
Inside the while loop, line~10 and line~11 has the highest complexity $O((s^t k^t)^d s^t k^t n)$ 
for the same reason. 
And let $n$ be the size of the initial CCA. 
As a result, the algorithm's time complexity is $O((s^t k^t)^d s^t k^t n^2)$. 
When $s, t$, and $d$ are fixed, the complexity is polynomial in $k$ and $n$.

%% file: section5.tex
\section{Experiments}\label{sec:exp}

In this section we show the results of experiments to evaluate the two proposed algorithms presented in the previous section. 
We focus on generation of $(1, 2)$-CDAs ($d=1$, $t=2$) for the following reasons.  
First, by nature of CDAs no interactions can be erroneously identified as non-faulty even when more than $d$ interactions are faulty; 
thus it is natural to set a small value to $d$ in practice.  
Second, the most common form of CIT targets two-way interactions (this form of CIT is called \emph{pair-wise testing}.)

\subsection{Experiment settings}

\begin{table}
    \centering
    \caption{Benchmark Information}
    \label{tab:benchmarks}
    \input{benchmarks.tex}
\end{table}

We wrote C++ programs that implement the two algorithms. 
Our implementation~\cite{DBLP:journals/corr/abs-1907-01779} of the IPOG algorithm~\cite{IPOG} was used as a CCA generator for both algorithms, 
while the Z3 solver~(version 4.8.1) was used in the satisfiability-based algorithms. 
We performed experiments with a total of 27 benchmark instances, numbered from 1 to 27. 
Benchmarks No.~1 to 7 are taken from \cite{Gargantini:citlab:iwct2012} which are provided as part of the CitLab tool. 
Benchmarks No.~8 to 27 can be found in \cite{Segall2011}. 
The detailed information of these benchmark instances is shown in Table~\ref{tab:benchmarks}. 
In Table~\ref{tab:benchmarks}, the columns labeled with $|\mathcal{F}|$ and $|\phi|$ show the number of parameters and the number of constraints
($\phi$ is the conjunction of the constraints). 
Columns $|\mathcal{VI}_2|$ and $|\mathcal{I}_2\backslash\mathcal{VI}_2|$ show respectively the number of valid interactions and the number of invalid interactions. 
The last column labeled $|\mathcal{T}\succ T|$ shows the number of pairs of an interaction set $\mathcal{T}$ and an interaction $T$ such that $\mathcal{T}$ masks $T$. 
For instance, the first line in the table shows that the benchmark \textit{car} has 9 parameters with 15 constraints. In the test space there are 102 valid interactions and 42 invalid interactions. Among the valid interactions, there are 1,487 pairs of an interaction set and an interaction such that the interaction set masks the interaction. 
All experiments were conducted on a machine with Intel Core i7-8700 CPU, 64 GB memory and Ubuntu 18.04 LTS OS. 
For each benchmark instance, the two generation algorithms were executed five times. 
The timeout period for each run was set to 1800 seconds.

\subsection{Experimental results}


\begin{table}
    \centering
    \caption{Experimental results. Numbers with $*$ indicate that the algorithm did not terminate within the time limit, in which case the CDAs obtained are not necessarily minimum.}
    \centerline{
        \input{results-new.tex}
    }
    \label{tab:results}
\end{table}

The results of the experiments are summarized in Table~\ref{tab:results}. 
The leftmost column shows the benchmark IDs. 
The rest of the table is divided into two parts representing the results of generation time and the results of sizes of the generated CDAs. 
Both parts have two sections describing the experiment results of the two proposed algorithms respectively. 
For each problem instance, the average value is reported for the satisfiability-based algorithm as it is deterministic, while the maximum, minimum, and average values are reported for the two-step heuristic generation algorithm.

The numbers with asterisk (*) in the satisfiability-based algorithm's columns show that the generation did not terminate within the time limit. 
Because the algorithm repeatedly generates CDAs with sizes varying until the minimum one is found, CDAs that are not optimal are constructed during the course of execution. 
The values with asterisk (*) correspond to the smallest~(not necessarily optimal) CDAs that were obtained within the time limit. 
For example, for benchmark No.~3, the algorithm took 806.21 seconds to generate a CDA of size~28. 
However, when it was trying to generate a CDA of size~27, the run of the algorithm exceeded the 1800 second time limit. 
There are also some benchmark instances that the algorithm did not find even one CDA within the time limit. 
We use the symbol ``--'' to indicate such a case. 
To compare the average consumed time of the two algorithms, the better results~(i.e., the shorter time) are denoted in bold font. 
The smaller average sizes of generated CDAs are also denoted in bold font. 

The satisfiability-based algorithm completed the generation process for three instances, namely, No.~1, 6, and 11. 
The CDAs obtained for these instances are all optimal.  
The algorithm was able to find small CDAs for some remaining instances (though it timed out), whereas it failed to find even a single CDA for others. 
In contrast, the two-step algorithms successfully generated CDAs for all benchmark instances. 
In addition, the execution time of the satisfiability-based algorithm was always much longer than the other algorithm, sometimes three orders of magnitude longer. 
There are two main reasons why the satisfiability-based algorithm is so slow. 
One reason is that the algorithm generates multiple CDAs in a single run. 
As stated in Section~\ref{sec:algo}, it generates $(d,t)$-CDAs with sizes varying from the size of a $(d+t)$-CCA. 
The CCA's size simply serves as the upper bound on the minimum CDA size: 
As it is not tight bound in general, to obtain an optimal $(d,t)$-CDA, the satisfiability solver is executed multiple times.  
The other reason, which is more obvious, is that satisfiability check may be time-consuming. 
The time required for the check becomes very long especially when the algorithm tries to find a CDA of minimum size minus 1, 
in which case the answer of the check is UNSAT (unsatisfiable). 
In the field of satisfiability, it is well known that UNSAT instances are usually more difficult than SAT instances. 


The satisfiability-based algorithm is deterministic. 
As stated above, the CCA size affects the algorithm's execution time and, if timeout occurs, the resulting CDA size. 
In contrast, the two-step heuristic algorithm is inherently nondeterministic: it generates different CDAs for different runs.  
The algorithm decreases the array size by repeatedly removing from the current array a test case selected at random~(line 7, Algorithm~\ref{alg:main}).
A test case can be removed only if the array remains to be a CDA after its removal; thus which test case is removed depends strongly on earlier selections.
Hence, different orders in which test cases are deleted lead to different CDAs. 

Another observation is that the two-step heuristic algorithm generated smaller CDAs than the satisfiability-based algorithm for No.~7. 
For the case, the satisfiability-based algorithm ran out of time before searching for minimum or near minimum CDAs. 
In view of these, we conclude that the two-step heuristic algorithm has balanced capabilities with respect to running time and CDA sizes it generates.  

%% file: benchmarks.tex
\begin{tabular}{c|c|ccccc}
\hline 
ID & SUT & $|\mathcal{F}|$ & $|\phi|$ & $|\mathcal{VI}_2|$ & $|\mathcal{I}_2\backslash\mathcal{VI}_2|$ & $|\mathcal{T}\succ T|$\tabularnewline
\hline 
1 & car & 9 & 15 & 102 & 42 & 1,487\tabularnewline
2 & graph\_product\_line & 20 & 45 & 499 & 261 & 37,212\tabularnewline
3 & real\_fm & 14 & 23 & 275 & 89 & 5,368\tabularnewline
4 & aircraft\_fm & 13 & 19 & 239 & 73 & 2,647\tabularnewline
5 & connector\_fm & 20 & 37 & 537 & 223 & 49,038\tabularnewline
6 & movies\_app\_fm & 13 & 23 & 211 & 101 & 4,968\tabularnewline
7 & stack\_fm & 17 & 28 & 465 & 79 & 6,399\tabularnewline
\hline 
8 & banking1 & 5 & 112 & 102 & 0 & 0\tabularnewline
9 & banking2 & 15 & 3 & 473 & 3 & 208\tabularnewline
10 & comm\_protocol & 11 & 128 & 285 & 35 & 2,177\tabularnewline
11 & concurrency & 5 & 7 & 36 & 4 & 130\tabularnewline
12 & healthcare1 & 10 & 21 & 361 & 8 & 512\tabularnewline
13 & healthcare2 & 12 & 25 & 466 & 1 & 124\tabularnewline
14 & healthcare3 & 29 & 31 & 3,092 & 59 & 8,700\tabularnewline
15 & healthcare4 & 35 & 22 & 5,707 & 38 & 3,359\tabularnewline
16 & insurance & 14 & 0 & 4,573 & 0 & 0\tabularnewline
17 & network\_mgmt & 9 & 20 & 1,228 & 20 & 189\tabularnewline
18 & processor\_comm1 & 15 & 13 & 1,058 & 13 & 1,510\tabularnewline
19 & processor\_comm2 & 25 & 125 & 2,525 & 854 & 35,156\tabularnewline
20 & services & 13 & 388 & 1,819 & 16 & 1,088\tabularnewline
21 & storage1 & 4 & 95 & 53 & 18 & 112\tabularnewline
22 & storage2 & 5 & 0 & 126 & 0 & 0\tabularnewline
23 & storage3 & 15 & 48 & 1,020 & 120 & 3,400\tabularnewline
24 & storage4 & 20 & 24 & 3,491 & 24 & 0\tabularnewline
25 & storage5 & 23 & 151 & 5,342 & 246 & 10,095\tabularnewline
26 & system\_mgmt & 10 & 17 & 310 & 14 & 825\tabularnewline
27 & telecom & 10 & 21 & 440 & 11 & 151\tabularnewline
\hline 
\end{tabular}

%% file: results-new.tex
\begin{tabular}{c|cccc|cccc}
    \hline 
    \multirow{3}{*}{ID} & \multicolumn{4}{c|}{Time (second)} & \multicolumn{4}{c}{Size} \\
    \cline{2-9}
        & \multicolumn{1}{c}{SMT} & \multicolumn{3}{c|}{Two-step} & \multicolumn{1}{c}{SMT} & \multicolumn{3}{c}{Two-step} \\
        &   avg.                &   max.    &   min.    &   avg.    &   avg.            &   max.    &   min.    &   avg.    \\
    \hline
    1   &   2.58                        &   0.12    &   0.07    &   \textbf{0.08}    &   \textbf{12}                        &   12      &   12      &   \textbf{12}      \\
    2   &   --                          &   0.18    &   0.13    &   \textbf{0.15}    &   --                                 &   26      &   26      &   \textbf{26}      \\
    3   &   806.21\textsuperscript{*}   &   0.16    &   0.11    &   \textbf{0.13}    &   \textbf{28}\textsuperscript{*}     &   31      &   30      &   30.40   \\
    4   &   268.75\textsuperscript{*}   &   0.10    &   0.09    &   \textbf{0.10}    &   \textbf{14}\textsuperscript{*}     &   18      &   18      &   18      \\
    5   &   88.06\textsuperscript{*}    &   0.17    &   0.12    &   \textbf{0.14}    &   \textbf{18}\textsuperscript{*}     &   18      &   18      &   \textbf{18}      \\
    6   &   6.41                        &   0.12    &   0.09    &   \textbf{0.10}    &   \textbf{9}                         &   10      &   10      &   10      \\
    7   &   357.50\textsuperscript{*}   &   0.18    &   0.13    &   \textbf{0.15}    &   34\textsuperscript{*}              &   33      &   33      &   \textbf{33}      \\
    \hline
    8   &   1557.89\textsuperscript{*}  &   0.15    &   0.10    &   \textbf{0.13}    &   \textbf{25}\textsuperscript{*}     &   37      &   36      &   36.40   \\
    9   &   --                          &   0.12    &   0.09    &   \textbf{0.11}    &   --                                 &   42      &   42      &   \textbf{42}      \\
    10  &   --                          &   0.21    &   0.15    &   \textbf{0.17}    &   --                                 &   50      &   47      &   \textbf{48.60}   \\
    11  &   0.364                       &   0.10    &   0.07    &   \textbf{0.08}    &   \textbf{8}                         &   8       &   8       &   \textbf{8}       \\
    12  &   --                          &   0.17    &   0.12    &   \textbf{0.14}    &   --                                 &   95      &   92      &   \textbf{94}      \\
    13  &   --                          &   0.20    &   0.16    &   \textbf{0.18}    &   --                                 &   60      &   57      &   \textbf{58.20}   \\
    14  &   --                          &   3.14    &   2.79    &   \textbf{2.95}    &   --                                 &   179     &   173     &   \textbf{176}     \\
    15  &   --                          &   19.87   &   18.73   &   \textbf{19.23}   &   --                                 &   234     &   220     &   \textbf{227.60}  \\
    16  &   --                          &   145.14  &   130.48  &   \textbf{134.31}  &   --                                 &   1997    &   1959    &   \textbf{1971.40} \\
    17  &   --                          &   1.56    &   1.48    &   \textbf{1.52}    &   --                                 &   405     &   394     &   \textbf{399.40}  \\
    18  &   --                          &   0.67    &   0.59    &   \textbf{0.61}    &   --                                 &   114     &   111     &   \textbf{112}     \\
    19  &   --                          &   3.16    &   2.94    &   \textbf{3.03}    &   --                                 &   122     &   118     &   \textbf{120.20}  \\
    20  &   --                          &   4.82    &   4.76    &   \textbf{4.79}    &   --                                 &   430     &   413     &   \textbf{422.40}  \\
    21  &   2.53\textsuperscript{*}     &   0.12    &   0.10    &   \textbf{0.10}    &   \textbf{25}\textsuperscript{*}     &   25      &   25      &   \textbf{25}      \\
    22  &   --                          &   0.08    &   0.07    &   \textbf{0.07}    &   --                                 &   51      &   45      &   \textbf{47.60}   \\
    23  &   --                          &   0.64    &   0.58    &   \textbf{0.61}    &   --                                 &   189     &   185     &   \textbf{187.60}  \\
    24  &   --                          &   16.28   &   15.61   &   \textbf{15.88}   &   --                                 &   517     &   500     &   \textbf{506.20}  \\
    25  &   --                          &   130.25  &   120.02  &   \textbf{124.86}  &   --                                 &   860     &   843     &   \textbf{851.60}  \\
    26  &   --                          &   0.12    &   0.09    &   \textbf{0.11}    &   --                                 &   53      &   49      &   \textbf{51.80}   \\
    27  &   --                          &   0.19    &   0.14    &   \textbf{0.16}    &   --                                 &   102     &   98      &   \textbf{99.8}    \\
    \hline  
\end{tabular}

%% file: threats.tex
\section{Threats to Validity}\label{sec:threats}

The experimental results about the two CDA generation algorithms showed
 that the scalability of the satisfiability-based algorithm is 
 substantially limited, especially when comparing to the heuristic algorithm. 
 This conclusion heavily relies on the performance of the satisfiability solver 
 used in the implementation. 
Although Z3, which we adopted in our implementation, is one of the best known and fastest SMT solvers, 
solvers of SMT or similar problems, such as SAT, have seen constant progress. 
Hence the difference between the two algorithms might narrow in near future. 

The problem instances used in the experiments are well-known and have been used in many other studies; However, they do not necessarily capture the characteristics of all real-world problems. 
Although we believe that the algorithms' qualitative properties observed in the experiments are likely to hold in general, there can be new problems for which they do not hold. 

%% file: section6.tex
\section{Related Work}\label{sec:relatedwork}

CIT has been widely used for many years. 
In the practice of CIT, constraint handling has always been a vital issue.
Surveys about constraint handling for CIT include~\cite{8102999,wu2019survey,8913600}.

DAs, as well as LAs, were first introduced by Colbourn and McClary in \cite{colbourn_locatingarray2008}. They analyzed the mathematical properties of these arrays. 
As \cite{colbourn_locatingarray2008}, most of the studies on DAs and LAs focus on their mathematical aspects~\cite{Colbourn2018,7562157,10.1007/978-3-319-94667-2_29,Shi2012,ShiDetecting2012,Lu2019}. 
The application to screening experiments for TCP throughput in a mobile wireless network was reported in~\cite{Aldaco:2015,7528941}.
Other types of arrays that are intended for fault location include
\emph{Error Locating Arrays}~\cite{ELA2010} and 
\emph{Consecutive Detecting Arrays}~\cite{Shi2020}.

The concept of CLAs was first introduced in~\cite{Jin2018}.
Later a computational construction algorithm was proposed in~\cite{8639696}. 
In~\cite{JIN2020110771} the results of applying CLAs to fault localization for real-world programs were reported.

The present paper extends our earlier works:~\cite{10.1145/3341105.3373952} and~\cite{9338632}. 
In~\cite{10.1145/3341105.3373952} we introduced $(d,t)$-CDAs for the first time, together 
with a construction algorithm using an SMT-solver. The present paper introduces the other variants of CDAs, clarifies their relations, and refines the algorithm. The two-step heuristic algorithm was first 
proposed in~\cite{9338632}. In the current paper, we improved the implementation of the algorithm and conducted a new set of experiments to compare the two different algorithms 
with the new implementations.  

There are many other approaches to faulty interaction localization without using CDAs or other related arrays. 
One of such approaches is the use of adaptive testing~\cite{Wang2010,Zhang2011,Li2012,8728941,8728924,10.1109/TSE.2018.2865772,Niu2013,Niu2021}. 
In adaptive testing, when a failure is encountered, new test cases are adaptively generated and executed to narrow down possible causes.
On the other hand, testing using CDAs is nonadaptive in the sense that test outcomes do not alter future test plans.  
A clear benefit of using nonadaptive testing is the execution of test suites, which is often the most time-consuming part of the whole testing process, can be parallelized. 

CDAs and other arrays of similar kinds are intended to provide sufficient test outcomes to uniquely identify faulty interactions. On the other hand, some studies attempt to infer faulty interactions from insufficient information with, for example, machine learning. 
The studies in this line include~\cite{Yilmaz2006,Shakya2012,Nishiura2017}. 

For other approaches to identification of faulty interactions, readers are referred to 
a recent survey~\cite{9156014}.

%% file: section7.tex
\section{Conclusion}\label{sec:con}

In this paper, we introduced the notion of Constrained Detecting Arrays~(CDAs), which incorporates constraints among test parameters into Detecting Arrays~(DAs). 
CDAs generalize DAs so that localization of faulty interactions can be performed for systems with constraints. 
We proved several properties of CDAs as well as those that relate CDAs with other array structures, such as Constrained Covering Arrays~(CCAs) and Constrained Locating Arrays~(CLAs). 
We then proposed two generation algorithms. 
The first algorithm generates optimal CDAs using an off-the-shelf satisfiability solver. 
The second algorithm is heuristic and generates near-optimal CDAs in a reasonable time.  
The experimental results of both algorithms indicated that the heuristic algorithm can scale to problems of practical sizes. 

There are several possible directions for future work.
One direction is to apply CDAs to the testing of real-world programs to identify faulty interactions. 
The development of new algorithms for CDA construction also deserves further study. 
We believe that both meta-heuristic search and greedy heuristics may be promising because 
they have proved to be useful for the construction of CCAs. 
A recent study attempts to provide a systematic framework to compare CCA generators~\cite{Andrea2021}.
Applying such a framework to compare different CDA construction algorithms is also of interest. 

%% file: main.bbl
\begin{thebibliography}{42}
\expandafter\ifx\csname natexlab\endcsname\relax\def\natexlab#1{#1}\fi
\providecommand{\url}[1]{\texttt{#1}}
\providecommand{\href}[2]{#2}
\providecommand{\path}[1]{#1}
\providecommand{\DOIprefix}{doi:}
\providecommand{\ArXivprefix}{arXiv:}
\providecommand{\URLprefix}{URL: }
\providecommand{\Pubmedprefix}{pmid:}
\providecommand{\doi}[1]{\href{http://dx.doi.org/#1}{\path{#1}}}
\providecommand{\Pubmed}[1]{\href{pmid:#1}{\path{#1}}}
\providecommand{\bibinfo}[2]{#2}
\ifx\xfnm\relax \def\xfnm[#1]{\unskip,\space#1}\fi
\bibitem[{Kuhn et~al.(2013)Kuhn, Kacker, and Lei}]{kuhn_introductionCT2013}
\bibinfo{author}{D.~R. Kuhn}, \bibinfo{author}{R.~N. Kacker},
  \bibinfo{author}{Y.~Lei}, \bibinfo{title}{Introduction to combinatorial
  testing}, \bibinfo{publisher}{CRC Press}, \bibinfo{year}{2013}.
\bibitem[{Kuhn and Wallace(2004)}]{kuhn_faultinteractions2004}
\bibinfo{author}{D.~R. Kuhn}, \bibinfo{author}{D.~R. Wallace},
\newblock \bibinfo{title}{Software fault interactions and implications for
  software testing},
\newblock \bibinfo{journal}{IEEE Transactions on Software Engineering}
  \bibinfo{volume}{30} (\bibinfo{year}{2004}) \bibinfo{pages}{418--421}.
\bibitem[{Colbourn and McClary(2008)}]{colbourn_locatingarray2008}
\bibinfo{author}{C.~J. Colbourn}, \bibinfo{author}{D.~W. McClary},
\newblock \bibinfo{title}{Locating and detecting arrays for interaction
  faults},
\newblock \bibinfo{journal}{Journal of Combinatorial Optimization}
  \bibinfo{volume}{15} (\bibinfo{year}{2008}) \bibinfo{pages}{17--48}.
\bibitem[{Lin et~al.(2015)Lin, Luo, Cai, Su, Hao, and Zhang}]{lin_TCA2015}
\bibinfo{author}{J.~Lin}, \bibinfo{author}{C.~Luo}, \bibinfo{author}{S.~Cai},
  \bibinfo{author}{K.~Su}, \bibinfo{author}{D.~Hao},
  \bibinfo{author}{L.~Zhang},
\newblock \bibinfo{title}{{TCA}: An efficient two-mode meta-heuristic algorithm
  for combinatorial test generation},
\newblock in: \bibinfo{booktitle}{Proc. of the 30th International Conference on
  Automated Software Engineering (ASE)}, \bibinfo{publisher}{ACM/IEEE},
  \bibinfo{year}{2015}, pp. \bibinfo{pages}{494--505}.
\bibitem[{Shiba et~al.(2004)Shiba, Tsuchiya, and Kikuno}]{Shiba:2004}
\bibinfo{author}{T.~Shiba}, \bibinfo{author}{T.~Tsuchiya},
  \bibinfo{author}{T.~Kikuno},
\newblock \bibinfo{title}{Using artificial life techniques to generate test
  cases for combinatorial testing},
\newblock in: \bibinfo{booktitle}{Proc. of 28th Annual International Computer
  Software and Applications Conference (COMPSAC '04)}, \bibinfo{year}{2004},
  pp. \bibinfo{pages}{71--77}.
\bibitem[{Lei et~al.(2008)Lei, Kacker, Kuhn, Okun, and Lawrence}]{IPOG}
\bibinfo{author}{Y.~Lei}, \bibinfo{author}{R.~Kacker}, \bibinfo{author}{D.~R.
  Kuhn}, \bibinfo{author}{V.~Okun}, \bibinfo{author}{J.~Lawrence},
\newblock \bibinfo{title}{Ipog/ipog-d: efficient test generation for multi-way
  combinatorial testing},
\newblock \bibinfo{journal}{Software Testing, Verification and Reliability}
  \bibinfo{volume}{18} (\bibinfo{year}{2008}) \bibinfo{pages}{125--148}.
\bibitem[{Jin and Tsuchiya(2020)}]{JIN2020110771}
\bibinfo{author}{H.~Jin}, \bibinfo{author}{T.~Tsuchiya},
\newblock \bibinfo{title}{Constrained locating arrays for combinatorial
  interaction testing},
\newblock \bibinfo{journal}{Journal of Systems and Software}
  \bibinfo{volume}{170} (\bibinfo{year}{2020}) \bibinfo{pages}{110771}.
\bibitem[{Hu et~al.(2020)Hu, Wong, Kuhn, and Kacker}]{HuESE2020}
\bibinfo{author}{L.~Hu}, \bibinfo{author}{W.~E. Wong}, \bibinfo{author}{D.~R.
  Kuhn}, \bibinfo{author}{R.~N. Kacker},
\newblock \bibinfo{title}{How does combinatorial testing perform in the real
  world: an empirical study},
\newblock \bibinfo{journal}{Empirical Software Engineering}
  \bibinfo{volume}{25} (\bibinfo{year}{2020}) \bibinfo{pages}{2661--2693}.
\bibitem[{Hnich et~al.(2006)Hnich, Prestwich, Selensky, and
  Smith}]{hnich_constraint2006}
\bibinfo{author}{B.~Hnich}, \bibinfo{author}{S.~D. Prestwich},
  \bibinfo{author}{E.~Selensky}, \bibinfo{author}{B.~M. Smith},
\newblock \bibinfo{title}{Constraint models for the covering test problem},
\newblock \bibinfo{journal}{Constraints} \bibinfo{volume}{11}
  (\bibinfo{year}{2006}) \bibinfo{pages}{199--219}.
\bibitem[{Tsuchiya(2019)}]{DBLP:journals/corr/abs-1907-01779}
\bibinfo{author}{T.~Tsuchiya},
\newblock \bibinfo{title}{Using binary decision diagrams for constraint
  handling in combinatorial interaction testing},
\newblock \bibinfo{journal}{CoRR} \bibinfo{volume}{abs/1907.01779}
  (\bibinfo{year}{2019}).
\bibitem[{{Gargantini} and {Vavassori}(2012)}]{Gargantini:citlab:iwct2012}
\bibinfo{author}{A.~{Gargantini}}, \bibinfo{author}{P.~{Vavassori}},
\newblock \bibinfo{title}{{CitLab}: A laboratory for combinatorial interaction
  testing},
\newblock in: \bibinfo{booktitle}{2012 IEEE Fifth International Conference on
  Software Testing, Verification and Validation}, \bibinfo{year}{2012}, pp.
  \bibinfo{pages}{559--568}. \DOIprefix\doi{10.1109/ICST.2012.141}.
\bibitem[{Segall et~al.(2011)Segall, Tzoref-Brill, and Farchi}]{Segall2011}
\bibinfo{author}{I.~Segall}, \bibinfo{author}{R.~Tzoref-Brill},
  \bibinfo{author}{E.~Farchi},
\newblock \bibinfo{title}{Using binary decision diagrams for combinatorial test
  design},
\newblock in: \bibinfo{booktitle}{Proc. of the 2011 International Symposium on
  Software Testing and Analysis (ISSTA)}, \bibinfo{publisher}{ACM},
  \bibinfo{year}{2011}, pp. \bibinfo{pages}{254--264}.
\bibitem[{Ahmed et~al.(2017)Ahmed, Zamli, Afzal, and Bures}]{8102999}
\bibinfo{author}{B.~S. Ahmed}, \bibinfo{author}{K.~Z. Zamli},
  \bibinfo{author}{W.~Afzal}, \bibinfo{author}{M.~Bures},
\newblock \bibinfo{title}{Constrained interaction testing: A systematic
  literature study},
\newblock \bibinfo{journal}{IEEE Access} \bibinfo{volume}{5}
  (\bibinfo{year}{2017}) \bibinfo{pages}{1--1}.
\bibitem[{Wu et~al.(2019{\natexlab{a}})Wu, Nie, Petke, Jia, and
  Harman}]{wu2019survey}
\bibinfo{author}{H.~Wu}, \bibinfo{author}{C.~Nie}, \bibinfo{author}{J.~Petke},
  \bibinfo{author}{Y.~Jia}, \bibinfo{author}{M.~Harman}, \bibinfo{title}{A
  survey of constrained combinatorial testing},
  \bibinfo{year}{2019}{\natexlab{a}}.
  \href{http://arxiv.org/abs/1908.02480}{\tt arXiv:1908.02480}.
\bibitem[{Wu et~al.(2019{\natexlab{b}})Wu, Changhai, Petke, Jia, and
  Harman}]{8913600}
\bibinfo{author}{H.~Wu}, \bibinfo{author}{N.~Changhai},
  \bibinfo{author}{J.~Petke}, \bibinfo{author}{Y.~Jia},
  \bibinfo{author}{M.~Harman},
\newblock \bibinfo{title}{Comparative analysis of constraint handling
  techniques for constrained combinatorial testing},
\newblock \bibinfo{journal}{IEEE Transactions on Software Engineering}
  (\bibinfo{year}{2019}{\natexlab{b}}) \bibinfo{pages}{1--1}.
\bibitem[{Colbourn and Syrotiuk(2018)}]{Colbourn2018}
\bibinfo{author}{C.~J. Colbourn}, \bibinfo{author}{V.~R. Syrotiuk},
\newblock \bibinfo{title}{On a combinatorial framework for fault
  characterization},
\newblock \bibinfo{journal}{Mathematics in Computer Science}
  \bibinfo{volume}{12} (\bibinfo{year}{2018}) \bibinfo{pages}{429--451}.
\bibitem[{{Compton} et~al.(2016){Compton}, {Mehari}, {Colbourn}, {De Poorter},
  and {Syrotiuk}}]{7562157}
\bibinfo{author}{R.~{Compton}}, \bibinfo{author}{M.~T. {Mehari}},
  \bibinfo{author}{C.~J. {Colbourn}}, \bibinfo{author}{E.~{De Poorter}},
  \bibinfo{author}{V.~R. {Syrotiuk}},
\newblock \bibinfo{title}{Screening interacting factors in a wireless network
  testbed using locating arrays},
\newblock in: \bibinfo{booktitle}{2016 IEEE Conference on Computer
  Communications Workshops (INFOCOM WKSHPS)}, \bibinfo{year}{2016}, pp.
  \bibinfo{pages}{650--655}. \DOIprefix\doi{10.1109/INFCOMW.2016.7562157}.
\bibitem[{Seidel et~al.(2018)Seidel, Sarkar, Colbourn, and
  Syrotiuk}]{10.1007/978-3-319-94667-2_29}
\bibinfo{author}{S.~A. Seidel}, \bibinfo{author}{K.~Sarkar},
  \bibinfo{author}{C.~J. Colbourn}, \bibinfo{author}{V.~R. Syrotiuk},
\newblock \bibinfo{title}{Separating interaction effects using locating and
  detecting arrays},
\newblock in: \bibinfo{editor}{C.~Iliopoulos}, \bibinfo{editor}{H.~W. Leong},
  \bibinfo{editor}{W.-K. Sung} (Eds.), \bibinfo{booktitle}{Combinatorial
  Algorithms}, \bibinfo{publisher}{Springer International Publishing},
  \bibinfo{address}{Cham}, \bibinfo{year}{2018}, pp. \bibinfo{pages}{349--360}.
\bibitem[{Shi et~al.(2012{\natexlab{a}})Shi, Tang, and Yin}]{Shi2012}
\bibinfo{author}{C.~Shi}, \bibinfo{author}{Y.~Tang}, \bibinfo{author}{J.~Yin},
\newblock \bibinfo{title}{Optimal locating arrays for at most two faults},
\newblock \bibinfo{journal}{Science China Mathematics} \bibinfo{volume}{55}
  (\bibinfo{year}{2012}{\natexlab{a}}) \bibinfo{pages}{197--206}.
\bibitem[{Shi et~al.(2012{\natexlab{b}})Shi, Tang, and Yin}]{ShiDetecting2012}
\bibinfo{author}{C.~Shi}, \bibinfo{author}{Y.~Tang}, \bibinfo{author}{J.~Yin},
\newblock \bibinfo{title}{The equivalence between optimal detecting arrays and
  super-simple {OA}s},
\newblock \bibinfo{journal}{Designs, Codes and Cryptography}
  \bibinfo{volume}{62} (\bibinfo{year}{2012}{\natexlab{b}})
  \bibinfo{pages}{131--142}.
\bibitem[{Lu and Jimbo(2019)}]{Lu2019}
\bibinfo{author}{X.-N. Lu}, \bibinfo{author}{M.~Jimbo},
\newblock \bibinfo{title}{Arrays for combinatorial interaction testing: a
  review on constructive approaches},
\newblock \bibinfo{journal}{Japanese Journal of Statistics and Data Science}
  \bibinfo{volume}{2} (\bibinfo{year}{2019}) \bibinfo{pages}{641--667}.
\bibitem[{Aldaco et~al.(2015)Aldaco, Colbourn, and Syrotiuk}]{Aldaco:2015}
\bibinfo{author}{A.~N. Aldaco}, \bibinfo{author}{C.~J. Colbourn},
  \bibinfo{author}{V.~R. Syrotiuk},
\newblock \bibinfo{title}{Locating arrays: {A} new experimental design for
  screening complex engineered systems},
\newblock \bibinfo{journal}{SIGOPS Oper. Syst. Rev.} \bibinfo{volume}{49}
  (\bibinfo{year}{2015}) \bibinfo{pages}{31--40}.
\bibitem[{{Colbourn} and {Syrotiuk}(2016)}]{7528941}
\bibinfo{author}{C.~J. {Colbourn}}, \bibinfo{author}{V.~R. {Syrotiuk}},
\newblock \bibinfo{title}{Coverage, location, detection, and measurement},
\newblock in: \bibinfo{booktitle}{2016 IEEE Ninth International Conference on
  Software Testing, Verification and Validation Workshops (ICSTW)},
  \bibinfo{year}{2016}, pp. \bibinfo{pages}{19--25}.
  \DOIprefix\doi{10.1109/ICSTW.2016.38}.
\bibitem[{Mart\'inez et~al.(2010)Mart\'inez, Moura, Panario, and
  Stevens}]{ELA2010}
\bibinfo{author}{C.~Mart\'inez}, \bibinfo{author}{L.~Moura},
  \bibinfo{author}{D.~Panario}, \bibinfo{author}{B.~Stevens},
\newblock \bibinfo{title}{Locating errors using {ELAs}, covering arrays, and
  adaptive testing algorithms},
\newblock \bibinfo{journal}{SIAM Journal on Discrete Mathematics}
  \bibinfo{volume}{23} (\bibinfo{year}{2010}) \bibinfo{pages}{1776--1799}.
\bibitem[{Shi et~al.(2020)Shi, Jiang, and Tao}]{Shi2020}
\bibinfo{author}{C.~Shi}, \bibinfo{author}{L.~Jiang}, \bibinfo{author}{A.~Tao},
\newblock \bibinfo{title}{Consecutive detecting arrays for interaction faults},
\newblock \bibinfo{journal}{Graphs and Combinatorics} \bibinfo{volume}{36}
  (\bibinfo{year}{2020}) \bibinfo{pages}{1203--1218}.
\bibitem[{Jin et~al.(2018)Jin, Kitamura, Choi, and Tsuchiya}]{Jin2018}
\bibinfo{author}{H.~Jin}, \bibinfo{author}{T.~Kitamura}, \bibinfo{author}{E.-H.
  Choi}, \bibinfo{author}{T.~Tsuchiya},
\newblock \bibinfo{title}{A satisfiability-based approach to generation of
  constrained locating arrays},
\newblock in: \bibinfo{booktitle}{2018 IEEE International Conference on
  Software Testing, Verification and Validation Workshops},
  \bibinfo{year}{2018}, pp. \bibinfo{pages}{285--294}.
  \DOIprefix\doi{10.1109/ICSTW.2018.00062}.
\bibitem[{{Jin} and {Tsuchiya}(2018)}]{8639696}
\bibinfo{author}{H.~{Jin}}, \bibinfo{author}{T.~{Tsuchiya}},
\newblock \bibinfo{title}{Deriving fault locating test cases from constrained
  covering arrays},
\newblock in: \bibinfo{booktitle}{2018 IEEE 23rd Pacific Rim International
  Symposium on Dependable Computing (PRDC)}, \bibinfo{year}{2018}, pp.
  \bibinfo{pages}{233--240}. \DOIprefix\doi{10.1109/PRDC.2018.00044}.
\bibitem[{Jin et~al.(2020)Jin, Shi, and Tsuchiya}]{10.1145/3341105.3373952}
\bibinfo{author}{H.~Jin}, \bibinfo{author}{C.~Shi},
  \bibinfo{author}{T.~Tsuchiya},
\newblock \bibinfo{title}{Constrained detecting arrays for fault localization
  in combinatorial testing},
\newblock in: \bibinfo{booktitle}{Proceedings of the 35th Annual ACM Symposium
  on Applied Computing}, SAC '20, \bibinfo{publisher}{Association for Computing
  Machinery}, \bibinfo{address}{New York, NY, USA}, \bibinfo{year}{2020}, p.
  \bibinfo{pages}{1971–1978}. \URLprefix
  \url{https://doi.org/10.1145/3341105.3373952}.
  \DOIprefix\doi{10.1145/3341105.3373952}.
\bibitem[{Jin and Tsuchiya(2020)}]{9338632}
\bibinfo{author}{H.~Jin}, \bibinfo{author}{T.~Tsuchiya},
\newblock \bibinfo{title}{A two-step heuristic algorithm for generating
  constrained detecting arrays for combinatorial interaction testing},
\newblock in: \bibinfo{booktitle}{2020 IEEE 29th International Conference on
  Enabling Technologies: Infrastructure for Collaborative Enterprises
  (WETICE)}, \bibinfo{year}{2020}, pp. \bibinfo{pages}{219--224}.
  \DOIprefix\doi{10.1109/WETICE49692.2020.00050}.
\bibitem[{{Wang} et~al.(2010){Wang}, {Xu}, {Chen}, and {Xu}}]{Wang2010}
\bibinfo{author}{Z.~{Wang}}, \bibinfo{author}{B.~{Xu}},
  \bibinfo{author}{L.~{Chen}}, \bibinfo{author}{L.~{Xu}},
\newblock \bibinfo{title}{Adaptive interaction fault location based on
  combinatorial testing},
\newblock in: \bibinfo{booktitle}{2010 10th International Conference on Quality
  Software}, \bibinfo{year}{2010}, pp. \bibinfo{pages}{495--502}.
  \DOIprefix\doi{10.1109/QSIC.2010.36}.
\bibitem[{Zhang and Zhang(2011)}]{Zhang2011}
\bibinfo{author}{Z.~Zhang}, \bibinfo{author}{J.~Zhang},
\newblock \bibinfo{title}{Characterizing failure-causing parameter interactions
  by adaptive testing},
\newblock in: \bibinfo{booktitle}{Proceedings of the 20th International
  Symposium on Software Testing and Analysis, {ISSTA} 2011, Toronto, ON,
  Canada, July 17-21, 2011}, \bibinfo{year}{2011}, pp.
  \bibinfo{pages}{331--341}. \URLprefix
  \url{https://doi.org/10.1145/2001420.2001460}.
  \DOIprefix\doi{10.1145/2001420.2001460}.
\bibitem[{Li et~al.(2012)Li, Nie, and Lei}]{Li2012}
\bibinfo{author}{J.~Li}, \bibinfo{author}{C.~Nie}, \bibinfo{author}{Y.~Lei},
\newblock \bibinfo{title}{Improved delta debugging based on combinatorial
  testing},
\newblock in: \bibinfo{booktitle}{2012 12th International Conference on Quality
  Software, Xi'an, Shaanxi, China, August 27-29, 2012}, \bibinfo{year}{2012},
  pp. \bibinfo{pages}{102--105}. \URLprefix
  \url{https://doi.org/10.1109/QSIC.2012.28}.
  \DOIprefix\doi{10.1109/QSIC.2012.28}.
\bibitem[{Arcaini et~al.(2019)Arcaini, Gargantini, and Radavelli}]{8728941}
\bibinfo{author}{P.~Arcaini}, \bibinfo{author}{A.~Gargantini},
  \bibinfo{author}{M.~Radavelli},
\newblock \bibinfo{title}{Efficient and guaranteed detection of t-way
  failure-inducing combinations},
\newblock in: \bibinfo{booktitle}{2019 IEEE International Conference on
  Software Testing, Verification and Validation Workshops (ICSTW)},
  \bibinfo{year}{2019}, pp. \bibinfo{pages}{200--209}.
  \DOIprefix\doi{10.1109/ICSTW.2019.00054}.
\bibitem[{Bonn et~al.(2019)Bonn, Foegen, and Lichter}]{8728924}
\bibinfo{author}{J.~Bonn}, \bibinfo{author}{K.~Foegen},
  \bibinfo{author}{H.~Lichter},
\newblock \bibinfo{title}{A framework for automated combinatorial test
  generation, execution, and fault characterization},
\newblock in: \bibinfo{booktitle}{2019 IEEE International Conference on
  Software Testing, Verification and Validation Workshops (ICSTW)},
  \bibinfo{year}{2019}, pp. \bibinfo{pages}{224--233}.
  \DOIprefix\doi{10.1109/ICSTW.2019.00057}.
\bibitem[{Niu et~al.(2020)Niu, Nie, Leung, Lei, Wang, Xu, and
  Wang}]{10.1109/TSE.2018.2865772}
\bibinfo{author}{X.~Niu}, \bibinfo{author}{C.~Nie}, \bibinfo{author}{H.~Leung},
  \bibinfo{author}{Y.~Lei}, \bibinfo{author}{X.~Wang}, \bibinfo{author}{J.~Xu},
  \bibinfo{author}{Y.~Wang},
\newblock \bibinfo{title}{An interleaving approach to combinatorial testing and
  failure-inducing interaction identification},
\newblock \bibinfo{journal}{IEEE Trans. Softw. Eng.} \bibinfo{volume}{46}
  (\bibinfo{year}{2020}) \bibinfo{pages}{584–615}.
\bibitem[{{Niu} et~al.(2013){Niu}, {Nie}, {Lei}, and {Chan}}]{Niu2013}
\bibinfo{author}{X.~{Niu}}, \bibinfo{author}{C.~{Nie}},
  \bibinfo{author}{Y.~{Lei}}, \bibinfo{author}{A.~T.~S. {Chan}},
\newblock \bibinfo{title}{Identifying failure-inducing combinations using tuple
  relationship},
\newblock in: \bibinfo{booktitle}{2013 IEEE Sixth International Conference on
  Software Testing, Verification and Validation Workshops},
  \bibinfo{year}{2013}, pp. \bibinfo{pages}{271--280}.
  \DOIprefix\doi{10.1109/ICSTW.2013.38}.
\bibitem[{Niu et~al.(2021)Niu, Wu, Changhai, Lei, and Wang}]{Niu2021}
\bibinfo{author}{X.~Niu}, \bibinfo{author}{H.~Wu},
  \bibinfo{author}{N.~Changhai}, \bibinfo{author}{Y.~Lei},
  \bibinfo{author}{X.~Wang},
\newblock \bibinfo{title}{A theory of pending schemas in combinatorial
  testing},
\newblock \bibinfo{journal}{IEEE Transactions on Software Engineering}
  (\bibinfo{year}{2021}) \bibinfo{pages}{1--1}.
\bibitem[{Yilmaz et~al.(2006)Yilmaz, Cohen, and Porter}]{Yilmaz2006}
\bibinfo{author}{C.~Yilmaz}, \bibinfo{author}{M.~B. Cohen},
  \bibinfo{author}{A.~A. Porter},
\newblock \bibinfo{title}{Covering arrays for efficient fault characterization
  in complex configuration spaces},
\newblock \bibinfo{journal}{{IEEE} Trans. Software Eng.} \bibinfo{volume}{32}
  (\bibinfo{year}{2006}) \bibinfo{pages}{20--34}.
\bibitem[{Shakya et~al.(2012)Shakya, Xie, Li, Lei, Kacker, and
  Kuhn}]{Shakya2012}
\bibinfo{author}{K.~Shakya}, \bibinfo{author}{T.~Xie}, \bibinfo{author}{N.~Li},
  \bibinfo{author}{Y.~Lei}, \bibinfo{author}{R.~Kacker}, \bibinfo{author}{D.~R.
  Kuhn},
\newblock \bibinfo{title}{Isolating failure-inducing combinations in
  combinatorial testing using test augmentation and classification},
\newblock in: \bibinfo{booktitle}{Fifth {IEEE} International Conference on
  Software Testing, Verification and Validation, {ICST} 2012, Montreal, QC,
  Canada, April 17-21, 2012}, \bibinfo{year}{2012}, pp.
  \bibinfo{pages}{620--623}. \URLprefix
  \url{https://doi.org/10.1109/ICST.2012.149}.
  \DOIprefix\doi{10.1109/ICST.2012.149}.
\bibitem[{{Nishiura} et~al.(2017){Nishiura}, {Choi}, and
  {Mizuno}}]{Nishiura2017}
\bibinfo{author}{K.~{Nishiura}}, \bibinfo{author}{E.~{Choi}},
  \bibinfo{author}{O.~{Mizuno}},
\newblock \bibinfo{title}{Improving faulty interaction localization using
  logistic regression},
\newblock in: \bibinfo{booktitle}{2017 IEEE International Conference on
  Software Quality, Reliability and Security (QRS)}, \bibinfo{year}{2017}, pp.
  \bibinfo{pages}{138--149}. \DOIprefix\doi{10.1109/QRS.2017.24}.
\bibitem[{Friedrichs et~al.(2020)Friedrichs, F\"ogen, and Lichter}]{9156014}
\bibinfo{author}{T.~Friedrichs}, \bibinfo{author}{K.~F\"ogen},
  \bibinfo{author}{H.~Lichter},
\newblock \bibinfo{title}{A comparison infrastructure for fault
  characterization algorithms},
\newblock in: \bibinfo{booktitle}{2020 IEEE International Conference on
  Software Testing, Verification and Validation Workshops (ICSTW)},
  \bibinfo{year}{2020}, pp. \bibinfo{pages}{201--210}.
  \DOIprefix\doi{10.1109/ICSTW50294.2020.00042}.
\bibitem[{Bombarda et~al.(2021)Bombarda, Crippa, and Gargantini}]{Andrea2021}
\bibinfo{author}{A.~Bombarda}, \bibinfo{author}{E.~Crippa},
  \bibinfo{author}{A.~Gargantini},
\newblock \bibinfo{title}{An environment for benchmarking combinatorial test
  suite generators},
\newblock in: \bibinfo{booktitle}{2021 IEEE International Conference on
  Software Testing, Verification and Validation Workshops (ICSTW)},
  \bibinfo{year}{2021}, pp. \bibinfo{pages}{48--56}.
  \DOIprefix\doi{10.1109/ICSTW52544.2021.00021}.

\end{thebibliography}
